\documentclass[aps,11pt,prd,superscriptaddress,onecolumn,nofootinbib,english,longbibliography,preprintnumbers,hidelinks]{revtex4}
\usepackage{amsmath}
\usepackage{amssymb}
\usepackage{graphicx}
\usepackage{orcidlink}
\usepackage{slashed}
\usepackage{xcolor}
\graphicspath{{plots}}
\usepackage{ragged2e} 
\usepackage{wrapfig}

\newcommand{\dd}{\text{d}}
\newcommand{\parent}[1]{\left(#1\right)}
\newcommand{\pIR}{\phi_{\text{\tiny IR}}}

\newcommand{\pM}{\phi_\text{\tiny M}}
\newcommand{\pQ}{\phi_\text{\tiny Q}}
\newcommand{\ps}{\phi_s}
\newcommand{\pv}{\phi_v}

\newcommand{\jp}{\varphi}
\newcommand{\per}{\ell}
\newcommand{\zs}{z_0}
\newcommand{\fs}{f_\text{\tiny IR}}

\newcommand{\psIR}{\phi_\text{\tiny IR}}
\newcommand{\gs}{g_\text{\tiny IR}}
\newcommand{\OO}{\mathcal{O}}

\newcommand{\fQFT}{f_{\text{\tiny QFT}}}

\newcommand{\e}{\epsilon}

\newcommand{\beq}{\begin{equation}}
\newcommand{\eeq}{\end{equation}}
\newcommand{\beqs}{\begin{eqnarray}}
\newcommand{\eeqs}{\end{eqnarray}}

\renewcommand{\L}{{\cal L}}

\renewcommand*{\a}{\alpha}

\renewcommand*{\d}{\delta}
\newcommand*{\h}{\eta}
\renewcommand*{\l}{\lambda}
\newcommand*{\m}{\mu}
\newcommand*{\n}{\nu}
\newcommand*{\f}{\phi}
\newcommand*{\F}{\Phi}
\renewcommand*{\r}{\rho}
\newcommand*{\y}{\psi}
\newcommand*{\s}{\sigma}
\newcommand*{\diff}{\text{d}}
\newcommand*{\p}{\partial}
\renewcommand*{\le}{\left}
\newcommand*{\ri}{\right}
\newcommand*{\cO}{\mathcal{O}}
\newcommand*{\gn}{G_\mathrm{N}}
\newcommand{\Vf}{V_5}

\newcommand{\spoIRi}{\mathfrak{a}^1_{(\text{\tiny IR})}}
\newcommand{\spoIRii}{\mathfrak{a}^{2}_{(\text{\tiny IR})}}
\newcommand{\spooUVi}{\mathfrak{a}^1_{\text{\tiny (0)}}}
\newcommand{\spofUVi}{\mathfrak{a}^1_{\text{\tiny (4)}}}
\newcommand{\spooUVii}{\mathfrak{a}^2_{\text{\tiny (0)}}}
\newcommand{\spotUVii}{\mathfrak{a}^2_{\text{\tiny (2)}}}
\newcommand{\spt}{\mathfrak{e}_{\mu\nu}^{\text{\tiny{(IR)}}}}
\newcommand{\sptUVo}{\mathfrak{e}_{\mu\nu}^{(0)}}
\newcommand{\sptUVf}{\mathfrak{e}_{\mu\nu}^{(4)}}

\newcommand{\sone}{s_1}
\newcommand{\vone}{v_1}
\newcommand{\stwo}{s_2}
\newcommand{\vtwo}{v_2}
\newcommand{\cone}{\mathsf{c}_1}
\newcommand{\ctwo}{\mathsf{c}_2}

\usepackage{ dsfont }

\newcommand{\orcidauthorFAEDO}{0000-0002-3887-2088} 
\newcommand{\orcidauthorHOYOS}{0000-0001-7612-0971} 
\newcommand{\orcidauthorPIAI}{0000-0002-2251-0111} 
\newcommand{\orcidauthorRODGERS}{0000-0002-4826-6540} 
\newcommand{\orcidauthorSUBILS}{0000-0003-0104-9722} 

\begin{document}

\begin{flushright}
			\hfill{NORDITA 2024-019}
\end{flushright}

\title{Light holographic dilatons near  critical points}

\author{Ant\'{o}n F. Faedo\,\orcidlink{\orcidauthorFAEDO}}
\email{anton.faedo@uniovi.es}
\affiliation{Departamento de F\'{i}sica, Universidad de Oviedo,  c/ Leopoldo Calvo Sotelo 18, ES-33007, Oviedo, Spain.}
\affiliation{Instituto Universitario de Ciencias y Tecnolog\'{\i}as Espaciales de Asturias (ICTEA), Calle de la Independencia 13, ES-33004, Oviedo, Spain.}

\author{Carlos Hoyos\,\orcidlink{\orcidauthorHOYOS}}
\email{hoyoscarlos@uniovi.es}
\affiliation{Departamento de F\'{i}sica, Universidad de Oviedo,  c/ Leopoldo Calvo Sotelo 18, ES-33007, Oviedo, Spain.}
\affiliation{Instituto Universitario de Ciencias y Tecnolog\'{\i}as Espaciales de Asturias (ICTEA), Calle de la Independencia 13, ES-33004, Oviedo, Spain.}

\author{Maurizio~Piai\,\orcidlink{\orcidauthorPIAI}}
\email{m.piai@swansea.ac.uk}
\affiliation{Department of Physics, Faculty  of Science and Engineering, Swansea University, Singleton Park, SA2 8PP, Swansea, Wales, UK}

\author{Ronnie Rodgers\,\orcidlink{\orcidauthorRODGERS}}
\email{ronnie.rodgers@su.se}
\affiliation{Nordita, Stockholm University and KTH Royal Institute of Technology,
	Hannes Alfvéns väg 12, SE-106 91 Stockholm, Sweden.}

\author{Javier~G.~Subils\,\orcidlink{\orcidauthorSUBILS}}
\email{j.gomezsubils@uu.nl}
\affiliation {Institute for Theoretical Physics, Utrecht University, 3584 CC Utrecht, The Netherlands}

\date{\today}

\begin{abstract}

We investigate the relation between the emergence of a dilaton in gapped (confining) field theories, and the presence of either complex fixed points or instabilities in the strongly coupled dynamics in two classes of bottom-up  holographic models.  We demonstrate that in one of the two classes there is a critical line of first-order phase transitions (at zero temperature) that terminates at a critical point. We calculate the mass spectrum of fluctuations of the associated regular  gravity backgrounds, which we interpret as bound states in the dual field theories. In proximity to the  second-order phase transition, we find a parametrically light scalar state, and its composition leads us to identify it as a dilaton.

\end{abstract}
\maketitle
\newpage
\tableofcontents


\section{Introduction}

The dilaton is the particle associated with the spontaneous breaking of approximate scale invariance~\cite{Coleman:1985rnk}. The  Higgs boson itself~\cite{ATLAS:2012yve,CMS:2012qbp}   is a prominent example of an approximate dilaton. Its couplings, which determine production cross-sections and decay rates, all  descend from this fundamental, symmetry related, property. A composite dilaton has been advocated  to  also emerge in special strongly coupled field theories~\cite{Migdal:1982jp}, in particular in the context of certain models of dynamical electroweak symmetry breaking (EWSB)~\cite{Leung:1985sn,Bardeen:1985sm,Yamawaki:1985zg,Holdom:1984sk}. The problem of establishing which strongly coupled theories  admit a dilaton in their composite spectrum has important implications, both  theoretical and phenomenological in nature~\cite{Goldberger:2007zk}, and has inspired a flurry of studies~\cite{Hong:2004td,Dietrich:2005jn,Vecchi:2010gj,Hashimoto:2010nw,DelDebbio:2021xwu,Zwicky:2023fay,Zwicky:2023krx}, particularly in view of the Large Hadron Collider (LHC) program~\cite{Eichten:2012qb,Elander:2012fk,Chacko:2012sy,Bellazzini:2012vz,Abe:2012eu,Bellazzini:2013fga,Hernandez-Leon:2017kea}. Yet, given the inherent difficulty of treating strongly coupled theories, there is no clear consensus in the field theory literature on the emergence of a dilaton as a composite state (see, for instance, the arguments in Refs.~\cite{Holdom:1986ub,Holdom:1987yu,Appelquist:2010gy,Grinstein:2011dq}).

In a parallel development,  the lattice community has uncovered evidence  of an anomalously light scalar particle, a flavor singlet, in the spectrum of  composite states in various theories~\cite{LatKMI:2014xoh,Appelquist:2016viq,LatKMI:2016xxi,Gasbarro:2017fmi,LatticeStrongDynamics:2018hun,LatticeStrongDynamicsLSD:2021gmp,Hasenfratz:2022qan,LSD:2023uzj,Fodor:2012ty,Fodor:2015vwa,Fodor:2016pls,Fodor:2017nlp,Fodor:2019vmw,Fodor:2020niv}. In all these cases, it is tempting to identify such a light state with  the dilaton. This finding has renewed interest in the study of dilaton effective field theory (dEFT), the low-energy description of the dilaton coupled to the pseudo-Nambu--Goldstone bosons (PNGBs) of the theory~\cite{Matsuzaki:2013eva,Golterman:2016lsd,Kasai:2016ifi,Hansen:2016fri,Golterman:2016cdd,Appelquist:2017wcg,Appelquist:2017vyy,Cata:2018wzl,Golterman:2018mfm,Cata:2019edh,Appelquist:2019lgk,Golterman:2020tdq,Golterman:2020utm, Appelquist:2022mjb}. Possible phenomenological applications of dEFT range from new dark matter models~\cite{Appelquist:2024koa} to composite Higgs models in which the Higgs boson is one of the PNGBs~\cite{Kaplan:1983fs,Georgi:1984af,Dugan:1984hq},\footnote{See the reviews in Refs.~\cite{Contino:2010rs,Panico:2015jxa,Witzel:2019jbe,Cacciapaglia:2020kgq,Bennett:2023wjw} and the useful  summary tables in Refs.~\cite{Ferretti:2013kya,Ferretti:2016upr,Cacciapaglia:2019bqz}.} as highlighted in Refs.~\cite{Appelquist:2020bqj,Appelquist:2022qgl}, by building on the ideas in Refs.~\cite{Vecchi:2015fma,Ma:2015gra,BuarqueFranzosi:2018eaj}.

The phrase {\it approximate scale invariance} is used to draw an analogy between the dilaton and the PNGBs associated with approximate continuous internal symmetries. In dEFT terms, the dilaton emerges when the spontaneous breaking of scale invariance dominates over its explicit breaking. Unfortunately, this requires tuning the dEFT, which in turns demands a dynamical explanation. Historically, the slowing down of the renormalization group (RG) flow (walking dynamics) in the underlying theory has been invoked as a microscopic explanation of such phenomena, already in the early papers on the subject~\cite{Leung:1985sn,Bardeen:1985sm,Yamawaki:1985zg}.

In this regard, a link between the elusive concept of approximate scale invariance and the classification of phase transitions in field theories was put forward in Ref.~\cite{Kaplan:2009kr}. There, walking dynamics was presented as a consequence of the presence of a conjugate pair of complex fixed points (FPs) in the vicinity of the RG flow of a theory. Later, in Refs.~\cite{Gorbenko:2018ncu,Gorbenko:2018dtm} it was proposed to think of the physics at the complex FPs as described by complex generalizations of  conformal field theories (CFTs). Within this approach, one aims at understanding properties of the real flow from the study of perturbations of the complex CFTs.

A major obstacle in dealing with these questions is the strongly coupled character of this phenomenon. It is therefore natural to address them using the gauge/gravity duality (also known as holography)~\cite{Maldacena:1997re,Gubser:1998bc,Witten:1998qj,Aharony:1999ti}, which provides an alternative framework to analyze field theories beyond their perturbative regime. In holography, the physics of special strong-coupling field theories can be understood in terms of classical gravity models living in higher dimensions. Already in Ref.~\cite{Kaplan:2009kr} it was argued that the instability of a gravity theory with approximately anti-de Sitter geometry coupled to fields with mass just below the  Breitenlohner--Freedman (BF) unitarity bound~\cite{Breitenlohner:1982jf} reproduces the physics of complex FPs, including Berezinskii--Kosterlitz--Thouless (BKT) phase transitions and walking dynamics. Concrete realizations of this mechanism in string theory exist, starting with Ref.~\cite{Jensen:2010ga}. On the other hand, a different holographic proposal that does not rely on violating the BF bound but on a realization of the gravity dual of the complex CFTs  through complex geometries was introduced in Ref.~\cite{Faedo:2019nxw}.\footnote{This approach can also lead to multiple hierarchies, see Ref.~\cite{Faedo:2021ksi}.}

In this paper we exploit the holographic approach to investigate the presence of a light dilaton in situations where strong-coupling effects lead to the emergence of a mass gap (e.g., in confining field theories). Confinement admits a geometric realization in gravity duals, in terms of  a portion of internal space smoothly shrinking to zero size~\cite{Witten:1998zw,Klebanov:2000hb, Maldacena:2000yy,Butti:2004pk}. Based on simple realizations of these ideas in models that are not derived from fundamental theories of gravity  (bottom-up holography), early phenomenological studies highlighted that under special conditions the spectrum contains a light state,  the radion~\cite{Goldberger:1999uk,DeWolfe:1999cp,
Goldberger:1999un,Csaki:2000zn,Arkani-Hamed:2000ijo,Rattazzi:2000hs,Kofman:2004tk}. This particle has the properties  of the dilaton. Yet, these earlier examples do not capture the dynamics of confinement in four dimensions, which historically  was argued to prevent the existence of a dilaton~\cite{Holdom:1986ub,Holdom:1987yu}. A plethora of studies addressed the question of whether it persists in more realistic holographic models~\cite{Elander:2011aa,Kutasov:2012uq,Evans:2013vca,Hoyos:2013gma,Megias:2014iwa,Elander:2015asa,Megias:2015qqh,Athenodorou:2016ndx,
Pomarol:2019aae,CruzRojas:2023jhw,Pomarol:2023xcc}, possibly within a fundamental theory~\cite{Elander:2009pk,Elander:2012yh,Elander:2013jqa,Elander:2017cle,Elander:2017hyr,Elander:2018aub,Elander:2018gte}. Refs.~\cite{Megias:2014iwa,Megias:2015qqh} showed that the existence of a light state is not a generic feature.

We propose and analyze a new class of  holographic models (dubbed {\it model A} later in the paper), in which the scalar potential admits at the same time a runaway direction that models confinement in the dual theory, along the lines of Ref.~\cite{Gursoy:2007cb,Gursoy:2007er}, as well as (complex) critical points---dual to complex CFTs. When these critical points are close to the real axis, the RG flow slows down when passing in between, giving rise to a hierarchy of scales with characteristic Miransky scaling \cite{Miransky:1984ef}, as detailed in \cite{Faedo:2019nxw}. The physics is quasi-conformal in this region, so it is natural to look for a light dilaton in the spectrum of the model. Varying the position of the complex CFTs, we find choices for which there is indeed a light scalar state. Confinement, hence, does not always lift the mass of the lightest state.

However, in the body of the paper, we will critically discuss the limitation of this class of models. It does not appear that the light scalar state found in the spectrum for backgrounds in model A is due to the presence of complex CFTs close to the renormalization group flow of the field theory, before leading to confinement. Rather, we find that a light scalar appears in regions of parameter space near special types of classical instabilities, at the boundaries of the phase space of  model A where their interpretation in field theory terms is uncertain. In order to expose the relation between these findings and the nature of possible phase transitions, we must build and analyze alternative models.

To this purpose, we build a class of regular backgrounds dubbed {\it model B}, in which confinement is modeled by the smooth shrinking to zero size of a portion of internal space~\cite{Witten:1998zw}. For the region of parameters relevant to our discussion, no instance of violation of the BF bound or proximity to complex CFTs is realised. Yet, these models possess classical instabilities in the gravity theory. Examples of one-parameter families of non-singular (super-)gravity solutions of this type have been found~\cite{Elander:2020ial,Elander:2020fmv,Elander:2021wkc}, even in simpler bottom-up  models~\cite{Elander:2022ebt}. In all of these, the mass of the lightest scalar in the spectrum continuously depends on a tunable parameter, which may be dialed to approach a tachyonic instability. However, a careful analysis reveals the presence of a first-order phase transition which prevents the approach to this instability. Hence, the scalar particle cannot be arbitrarily light in the physical region of parameter space.

Nevertheless, there is no known reason why  the first-order transition should be strong. We construct model B by borrowing the five-dimensional action from Ref.~\cite{Bea:2020ees}, which has been used for other purposes in Refs.~\cite{Bea:2021ieq, Ares:2020lbt,Bea:2021zol,Bea:2021zsu,Bea:2022mfb, Escriva:2022yaf}. Thanks to those studies, it is known that this model may lead to the required phase structure:  in the relevant region of parameter space, there is a line of first-order phase transitions that become weak, and eventually disappears into a crossover. At the end of the line, the transition is second-order and a critical point is encountered. As we shall see, in the vicinity of this critical point, the spectrum of bound states of the dual field theory contains a light scalar particle. We also present evidence to support the claim that this particle can be appropriately called a dilaton, which is the main new result of the study reported here. 

The paper is organized as follows. In Sec.~\ref{sec:fixed_point_annihilation} we present and analyze model A. We identify regions of parameter space in which a light scalar state emerges. These happen to be located in proximity to a boundary in parameter space, across which the nature of the gravity background solutions changes qualitatively. This is reminiscent of the findings in Refs.~\cite{Elander:2020ial,Elander:2020fmv,Elander:2021wkc}, and points to the existence of phase transitions, but we cannot determine their location and nature. We demonstrate that these findings are not directly related to the proximity of complex FPs to the renormalization group flow trajectories in the dual field theory.  In Sec.~\ref{Sec:B}, we introduce model B, within which regular confining geometries exist. A light state appears in the spectrum of fluctuations in the vicinity of a critical point bearing clear evidence of a second-order phase transition. We also show that the light scalar is an approximate dilaton, to which purpose we analyze its composition by comparing the calculation of the spectrum performed with and without including the effects of backreaction on the metric.  We summarize our results in the Discussion section, and relegate the more technical details to the appendices.

\section{Model A: soft-wall confinement and complex fixed points}
\label{sec:fixed_point_annihilation}

The first class of models we consider is defined by a one-scalar coupled to gravity in five dimensions, and by the choice of a scalar potential that leads to background solutions in which the long distance dynamics of the dual theory mimics a confining field theory---or, at least, a gapped theory. To this purpose, we borrow ideas from Ref.~\cite{Gursoy:2007er}, in order to make a representative choice of the potential. We then study the asymptotic behavior of the classical backgrounds, and present numerical examples of solutions that we interpret as providing the gravity dual description of such (putative) gapped field theories. We classify the properties of the solutions of interest, in particular the monotonicity of the background scalar field, and perform a local stability analysis by computing the spectrum of fluctuations.

\subsection{Models of type A}
\label{Sec:ModelA}

We consider gravity minimally coupled to a real scalar field, \(\f\), on a five-dimensional manifold, \(\mathcal{M}\), with action
\begin{equation} \label{eq:complex_fp_bulk_action}
    S = \frac{1}{16 \pi G_5} \int_{\mathcal{M}} \diff^5 x \sqrt{-g} \, \le[R - \frac{1}{2} g^{mn} \p_m \f \p_n \f - V_A(\f) \ri]
    + S_\mathrm{bdy}\,,
\end{equation}
where
\begin{equation} 
S_\mathrm{bdy} = \frac{1}{16\pi G_5}
\int_{\partial \mathcal{M}}\dd^4x \sqrt{-h} \, \left[2K+\frac{}{}4
\lambda_A(\phi)\right]\,.
\end{equation}
In these expressions, \(G_5\) is the gravitational constant in five dimensions, \(R\) is the Ricci scalar of the bulk metric \(g_{mn}\), and \(g\) is its determinant. We denote by \(S_\mathrm{bdy}\) boundary terms localized on $\partial \mathcal{M}$,  the (regularized) conformal  boundary of the asymptotically-\(AdS\) manifold, \(\mathcal{M}\). They are needed to render the on-shell action finite and the variational problem well-defined~\cite{deHaro:2000vlm}.  They include the Gibbons--Hawking--York  (GHY)  term, proportional to the extrinsic curvature, $K$, and a boundary-localized potential, $\lambda_A$. Throughout this paper, we  use the mostly-plus signature for the metric. In this section, Latin indices \(m,n,\dots\) denote five-dimensional bulk directions, while Greek indices \(\m,\n,\dots\) are used for directions on the four-dimensional boundary. 
The equations of motion following from the action in Eq.~\eqref{eq:complex_fp_bulk_action} are
\begin{gather}
    R_{mn} - \frac{1}{2} g_{mn} R = \frac{1}{2} \p_m \f \p_n \f - \frac{1}{2} g_{mn} \le[\frac{1}{2}(\p\f)^2 + V_A(\f) \ri],
    \qquad\quad
    \nabla^2 \f =  \p_\f V_A(\f).
    \label{eq:complex_fp_background_eom_general}
\end{gather}

In this section, we choose a scalar potential \(V_A\) that satisfies the relations
\begin{equation}\label{eq:dpot}
    \p_\f V_A(\f) = \frac{M^2}{\a} \le(\frac{\f^2}{\f_c^2} - 1\ri)\le(\frac{\f^2}{{\f_c^*}^2} - 1\ri)\sinh (\a \f), \qquad\quad V_A(0) = - \frac{12}{L^2}\,,
\end{equation}
where \(\f_c = \f_r + i \f_i\) is a complex constant (\(\f_{r,i} \in \mathbb{R}\)),  \(M^2\) and \(\a\) are real constants, and  \(L\) is the radius of curvature of the asymptotically-\(AdS\) solutions we will construct. Since Eq.~\eqref{eq:dpot} is independent of the sign of \(\a\), we will take \(\a>0\) without loss of generality. The potential enjoys a \(\mathbb{Z}_2\) symmetry under the transformation \(\f \to - \f\) and has a UV fixed point at \(\f=0\). Moreover, it has extrema at complex values of the field \(\f =\pm \f_c\) and \(\pm \f_c^*\), giving rise to complex AdS solutions that have been conjectured to be dual to complex CFTs \cite{Faedo:2019nxw}.\footnote{There are also complex extrema at \(\f = n \pi i /\a\) with \(n \in \mathbb{Z} \backslash \{0\}\). These will not play a role in the discussion.} It can be seen that, when these extrema are close to the real axis, that is, $\f_i\ll1$, the physics is approximately conformal and moreover an exponential hierarchy of scales is generated as dictated by Miransky scaling. 

By integrating \(\p_\f V_A(\f)\), we find the potential
\begin{eqnarray} \label{eq:complex_fp_scalar_potential}
    V_A(\f) &=& \frac{M^2 }{\a^2 |\f_c|^4} \biggl\{
        \le[\bigl(\f^2 - \f_c^2\bigr)\bigl(\f^2 - {\f_c^*}^2\bigr) + \frac{12 \f^2}{\a^2} - \frac{2}{\a^2}\bigl(\f_c^2 + {\f_c^*}^2\bigr)+ \frac{24}{\a^4} \ri] \cosh \le(\a\f \ri)
      \\   \nonumber
        && - \frac{2}{\a}   \le(2\f^2 - \f_c^2 - {\f_c^*}^2 + \frac{12}{\a^2} \ri) \f \sinh \le(\a \f\ri) + \frac{2}{\a^2} \bigl( \f_c^2 +  {\f_c^*}^2 \bigr) - \frac{24}{\a^4}
    \biggr\}
     - \frac{M^2}{\a^2}-\frac{12}{L^2}\,.
\end{eqnarray}
Expanding in powers of small \(\f\), one finds \(V_A(\f) = -12 L^{-2} + M^2 \f^2 / 2 + O(\f^3)\), hence we identify \(M^2\equiv-\Delta(4-\Delta)L^{-2}\) as the mass squared of the scalar field, which is related to the scaling dimension, $\Delta\leq 4$, of a scalar operator \(\cO\) in the dual quantum field theory. For large positive values of \(\f\), the potential diverges exponentially as
\begin{equation}
    V_A(\f) \approx \frac{M^2}{2\a^2|\f_c|^4} \f^4 e^{\a \f}  + \dots \,.
\end{equation}

We seek solutions to the equations of motion~\eqref{eq:complex_fp_background_eom_general} preserving Poincar\'e invariance in the dual field theory, a requirement that can be satisfied by adopting the following ansatz:
\begin{equation} \label{eq:complex_fp_background_ansatz}
    \diff s^2 = \frac{L^2}{z^2} \le[\diff z^2 + f(z) \, \h_{\m\n} \diff x^\m \, \diff x^\n\ri], \qquad \f = \f(z),
\end{equation}
where \(\h_{\m\n}\) is the four-dimensional Minkowski metric. The coordinates \(x^\m\) can be interpreted as the coordinates in the dual field theory, while \(z\) is the holographic radial coordinate, with \(z=0\) the conformal boundary. The non-trivial components of the equations of motion~\eqref{eq:complex_fp_background_eom_general} become
\begin{subequations} \label{eq:complex_fp_background_eom_ansatz}
\begin{align}
    z \frac{f'}{f} - \frac{z^2}{4} \frac{f'^2}{f^2} + \frac{z^2 \f'^2}{24} - \frac{L^2}{12}V_A - 1 &= 0,
    \label{eq:complex_fp_background_constraint}
    \\[0.5em]
    z^2 \frac{f''}{f} - 3 z \frac{f'}{f} + \frac{z^2 \f'^2}{6} + \frac{L^2}{3}V_A + 4 &= 0,
    \label{eq:complex_fp_background_metric_eom}
    \\[0.5em]
    z^2 \f'' + \le(2 z \frac{f'}{f} - 3 \ri) z \f' - L^2  \p_\f V_A&= 0,
    \label{eq:complex_fp_background_scalar_eom}
\end{align}
\end{subequations}
where primes indicate derivatives with respect to \(z\). These equations are not independent; equations~\eqref{eq:complex_fp_background_constraint} and~\eqref{eq:complex_fp_background_metric_eom} together imply~\eqref{eq:complex_fp_background_scalar_eom}. Note that the equations of motion are invariant under the rescaling \(f \to \Omega^2 f\) with constant \(\Omega\), corresponding to the freedom to rescale the field theory coordinates \(x^\m\). They are also invariant under the constant rescaling \(z \to \Omega z\).

The equations of motion~\eqref{eq:complex_fp_background_constraint} and~\eqref{eq:complex_fp_background_metric_eom} can be combined to yield
\begin{equation} \label{eq:complex_fp_eom_combination}
    z \le(\frac{z f'}{f}\ri)' + \frac{z^2 \f'^2}{3} = 0\,,
\end{equation}
which shows that 
\begin{equation} \label{eq:complex_fp_nec}
    \le(\frac{z f'}{f}\ri)' \leq 0\,,
\end{equation}
implying that \(z f'/f\) is  a monotonically decreasing function of \(z\).
This inequality can, equivalently, also be derived from the geometric form of the null energy condition. In the far UV (small \(z\)), we generically expect \(\f \simeq \Lambda z^{4-\Delta}\), where \(\Lambda\) is fixed by the boundary conditions and interpreted as the source for the scalar operator dual to \(\f\).
Solving Eq.~\eqref{eq:complex_fp_eom_combination} asymptotically at small \(z\), subject to the boundary condition \(f(0) = 1\), we then find that \(f(z) \approx 1 - \frac{\Lambda^2}{12} z^{2(4-\Delta)}\). The factor in brackets in Eq.~\eqref{eq:complex_fp_nec} is then
\begin{equation}
    \frac{z f'}{f} \simeq - \frac{(4-\Delta) \Lambda^2}{6} z^{2(4-\Delta)},
    \qquad
    (\mathrm{small}~z).
\end{equation}
Since this is negative, and since Eq.~\eqref{eq:complex_fp_nec} implies that \(z f'/f\) is a decreasing function of \(z\), we conclude that \(z f'/f\) is negative for \textit{all} values of  \(z>0\). Furthermore, since \(z\) and \(f\) are themselves non-negative, we conclude that \(f\) must be a decreasing function of \(z\),
\begin{equation} \label{eq:complex_fp_f_decreasing}
    f'(z) \leq 0.
\end{equation}

Finally, Eq.~\eqref{eq:complex_fp_background_constraint}, which is quadratic in \(f'/f\), admits one solution with \(f'/f \geq 0\) and another solution with \(f'/f \leq 0\). From Eq.~\eqref{eq:complex_fp_f_decreasing} one sees that the latter is the one consistent with  the asymptotically-\(AdS\)  boundary conditions. We can then substitute this solution for \(f'/f\) into Eq.~\eqref{eq:complex_fp_background_scalar_eom} to recast the equations of motion in nested form:
\begin{subequations}\label{eq:complex_fp_nested_eom}
\begin{align}
    z^2 \f''+ z \f' \le( 1 - 2\sqrt{ \frac{z^2 \f'^2 - 2 L^2 V}{6}}\ri) - L^2 \p_\f V &= 0,
    \label{eq:complex_fp_nested_eom_phi}
     \\
    z f' - \le(2 - \sqrt{ \frac{z^2 \f'^2 - 2 L^2 V}{6}} \ri)f &= 0.
    \label{eq:complex_fp_nested_eom_f}
\end{align}
\end{subequations}
This form of the equations of motion will be useful for constructing numerical solutions.

\subsection{Asymptotic behavior of solutions}
\label{sec:asyA}

As hinted at while introducing  model A, we will not consider all possible solutions of the background equations, for all possible choices of the parameters. Rather, we isolate a class of solutions that are representative of the gravity duals of field theories of interest. We hence find it useful to classify these solutions by discussing their asymptotic behavior,
in the region of the holographic direction, $z$, corresponding to the far ultraviolet (UV) and deep infrared  (IR) regimes in their dual field theory interpretation.

\subsubsection{Ultraviolet asymptotics}

From here on we  set the mass of the scalar field, \(\f\), such that \(M^2 L^2 = -3\), implying that the operator \(\cO\) dual to the field  \(\f\) has scaling dimension \(\Delta=3\). The equations of motion~\eqref{eq:complex_fp_background_eom_ansatz} then admit solutions which behave near the asymptotically-\(AdS_5\) boundary (at small \(z\)) as
\begin{subequations} \label{eq:complex_fp_background_uv}
\begin{align}
    f(z) &= 1 - \frac{\Lambda^2}{12} z^2  + O(z^4 \log z),
    \label{eq:complex_fp_background_uv_f}
    \\[1em]
    \f(z) &= \Lambda z + \le(2 - 3\a^2 + \frac{18}{\f_c^2} + \frac{18}{{\f_c^*}^2} \ri)\frac{\Lambda^3}{12} z^3 \log(|\Lambda| z) + \F z^3
    + O(z^5 \log z),
     \label{eq:complex_fp_background_uv_phi}
\end{align}
\end{subequations}
where \(\Lambda\) and \(\F\) are integration constants. We have used the freedom to rescale \(f\) in order to set \(f(0)=1\). All other coefficients in the expansions are fixed by the equations of motion. The constant \(\F\) determines the one-point function of the scalar operator \(\cO\) dual to \(\f\); as we show in appendix~\ref{app:complex_fp_holo_ren} we may choose a renormalization scheme in which 
\begin{equation}
\label{eq:complex_fp_Weyl_anomaly}
    \langle\cO\rangle = -\frac{a}{4} \F,
    \qquad
    a \equiv \frac{L^3}{2 \pi G_5}.
\end{equation}
Here, \(a\) is the coefficient of the Euler density appearing in the Weyl anomaly of the dual field theory~\cite{Henningson:1998gx},\footnote{If we place the field theory dual to our gravitational system on a curved spacetime, then the trace of its stress tensor is~\cite{Henningson:1998gx}
\[
    \langle T^\m{}_\m \rangle = - \frac{L^3}{2\pi G_5} \le(E_{(4)} + I_{(4)}\ri),
\]
where \(E_{(4)} = \frac{1}{64} \le(R^{\m\n\r\s}R_{\m\n\r\s} - 4 R^{\m\n} R_{\m\n} + R^2\ri)\) is the Euler density in four dimensions while \(I_{(4)}\) is proportional to the Weyl tensor contracted with itself, \(I_{(4)} = - \frac{1}{64} \le( R^{\m\n\r\s} R_{\m\n\r\s} - 2 R^{\m\n} R_{\m\n} + R^2/3 \ri) \).} and may thus be thought of as a measure of the number of degrees of freedom. In the same renormalization scheme, the renormalized on-shell action is
\begin{equation} \label{eq:complex_fp_renormalised_action}
    S_\mathrm{ren}^\star = \frac{a}{16}  \Lambda \F \, \mathrm{vol}(\mathbb{R}^{3,1}),
\end{equation}
where \(\mathrm{vol}(\mathbb{R}^{3,1})\) denotes the (regularized) volume of four-dimensional Minkowski space, arising from the integrals over the field theory directions.

\subsubsection{Infrared asymptotics}
\label{eq:asyAIR}

Inspired by Refs.~\cite{Gursoy:2007cb,Gursoy:2007er}, we seek solutions for which \(f(z)\) vanishes at some \(z=z_0\). To determine possible asymptotics near \(z=z_0\) it will be convenient to use the equations of motion in the nested form of Eqs.~\eqref{eq:complex_fp_nested_eom}. We require \(f'(z)/f(z) \to -\infty\) at \(z = z_0\),  hence Eq.~\eqref{eq:complex_fp_nested_eom_f} implies that we should seek singular solutions to Eq.~\eqref{eq:complex_fp_nested_eom_phi} that have \(\f \to \pm \infty\).\footnote{In principle, we could also look for solutions that have \(\f' \to \pm \infty\) but with \(\f\) remaining finite. We did not find any self-consistent asymptotic solutions to Eq.~\eqref{eq:complex_fp_nested_eom_phi} of this type.  
} We find that
\begin{align}
    e^{\pm \a \f(z)} &=  \frac{\a^2(8-3\a^2) |\f_c|^4}{36}  \frac{1}{\le(1 - z/z_0\ri)^{2} \le[ \log \le( 1 - z/z_0 \ri) \ri]^{4}}\,+\cdots\,,
    \nonumber \\
    f(z) & = f_\mathrm{IR} \le(1 - \frac{z}{z_0}\ri)^{4/3\a^2} \le|\log\le(1 - \frac{z}{z_0}\ri) \ri|^{16/3\a^2}\,+\cdots\,,
    \label{eq:complex_fp_ir_asymptotics}
\end{align}
where \(f_\mathrm{IR}\) is an integration constant, and the ellipses stand for terms that vanish for $z\rightarrow z_0$. The reality condition on $\phi$ implies that these solutions only exist for \(\a^2 < 8/3\). Depending on the sign chosen in the exponent of \(e^{\pm\a\f(z)}\), in Eq.~\eqref{eq:complex_fp_ir_asymptotics}, we find  that  \(\f \to \pm \infty\) as \(z \to z_0\). These two possibilities are related by the \(\mathbb{Z}_2\) symmetry of the potential in Eq.~\eqref{eq:complex_fp_scalar_potential}. 

Since we are interested in cases when the dual field theory exhibits a mass gap, we must restrict our attention to potentials with \(\a^2 \geq 2/3\).  This criterion~\cite{Gursoy:2007cb,Gursoy:2007er} is weaker than demonstrating confinement in the dual field theory. It can be derived by noting that the geometry defines a mass scale, \(\m\), through the relation~\cite{Csaki:2000cx}:
\begin{equation} ~\label{eq:geometric_mass_scale}
    \m^{-1} = 2 \int_0^{z_0} \frac{\diff z}{\sqrt{f(z)}}.
\end{equation}
Physically, \(\m^{-1}\) is the time it takes a massless particle to travel from a point at \(z=0\) to \(z = z_0\) and back, as measured by an observer at the conformal boundary at \(z=0\). When \(\a^2 < 2/3\), one finds  that \(f(z)\) vanishes faster than \((z_0 - z)^2\) as \(z \to z_0\), the integral on the right-hand side of Eq.~\eqref{eq:geometric_mass_scale} diverges, and hence \(\m\) is undefined. On the other hand, choosing  \(2/3 \leq \a^2 < 8/3\) leads to a finite \(\m\) that sets the scale of the mass gap.

We conclude this subsection with a digression. For completeness, we report here that we found also a second class of IR asymptotic solutions to the equations of motion, in which the leading-order behavior of the fields is
\begin{equation} \label{eq:complex_fp_ir_asymptotics_alt}
    e^{\f(z)} \approx \frac{e^{\f_\mathrm{IR}}}{ \le(1 - z/z_0\ri)^{\pm \sqrt{2/3}}}\,,
    \qquad
    f(z) \approx f_\mathrm{IR} \le(1 - \frac{z}{z_0}\ri)^{1/2}\,,
\end{equation}
where \(f_\mathrm{IR}\) and \(\f_\mathrm{IR}\) are integration constants. We disregard these solutions in the remainder of the paper since they exhibit a bad singularity by the criterion of Ref.~\cite{Gursoy:2008za}, namely the asymptotics of \(f(z)\) are such that an extra, IR boundary condition must be imposed to fix \(f_\mathrm{IR}\).

\subsection{Solutions}
\label{Sec:solA}

To construct the backgrounds of interest to this paper, we first construct asymptotic solutions to the equations of motion~\eqref{eq:complex_fp_nested_eom} near the end of space,  \(z=z_0\), with leading order behavior as in Eq.~\eqref{eq:complex_fp_ir_asymptotics}. Defining, for convenience, \(s=-\log(1-z/z_0)\), we find that such an asymptotic solution takes the following form:
\begin{subequations}
\begin{align}
    e^{\pm \a\f(z)} &= \frac{\a^2(8-3\a^2) |\f_c|^4}{36}   \frac{e^{2s}}{s^{4}} \sum_{n=0}^\infty \sum_{l=0}^n c^{(\f)}_{n,l}  \frac{(\log s)^l}{s^n},
    \label{eq:complex_fp_ir_expansion_phi}
    \\
    f(z) &= f_\mathrm{IR} e^{-4 s/3 \a^2} s^{16/3 \a^2}\sum_{n=0}^\infty \sum_{l=0}^n c^{(f)}_{n,l}  \frac{(\log s)^l}{s^n},
    \label{eq:complex_fp_ir_expansion_f}
\end{align}
\end{subequations}
where \(c^{(\f)}_{0,0} = c^{(f)}_{0,0} = 1\) and the remaining constant coefficients \(c^{(\f)}_{n,l}\) and \(c^{(f)}_{n,l}\) are determined iteratively by solving the equations of motion, having fixed $\phi_c$. We then proceed to construct our numerical background solutions, using the expansion in Eq.~\eqref{eq:complex_fp_ir_expansion_phi}---truncated at order \(n=6\)---to set boundary conditions for Eq.~\eqref{eq:complex_fp_nested_eom_phi} near \(z=z_0\) and then numerically integrating to small \(z\).
We fit the resulting numerical solution for \(\f(z)\) to the UV expansion (at small \(z\)) in Eq.~\eqref{eq:complex_fp_background_uv_phi},  and determine the UV coefficients \((\Lambda,\F)\) in units of \(z_0\). Substituting the solution for \(\f\) into Eq.~\eqref{eq:complex_fp_nested_eom_f}, we then numerically integrate equation~\eqref{eq:complex_fp_nested_eom_f} using the IR expansion~\eqref{eq:complex_fp_ir_expansion_f} with \(f_\mathrm{IR}=1\)  to set boundary conditions close to \(z=z_0\).  This generically gives rise to a solution with \(f(z=0) \neq 1\). Since Eq.~\eqref{eq:complex_fp_nested_eom_f} is linear in \(f\), we rescale the numerical solution for \(f(z)\) to obtain a solution with  \(f(z=0) = 1\), which finally takes the form in Eq.~\eqref{eq:complex_fp_background_uv_f}. In doing so we find that \(f_\mathrm{IR} \neq 1\), as desired.

\begin{figure}[t]
    \begin{center}
    \includegraphics{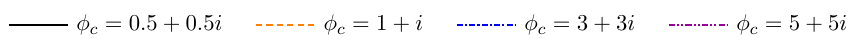}\\
    \includegraphics{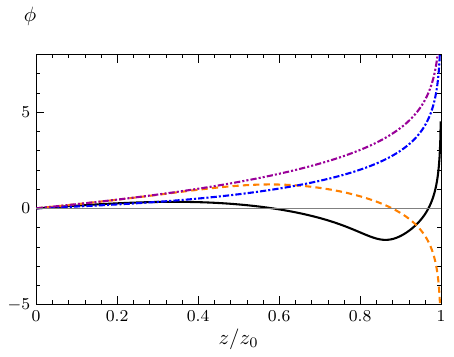}
    \includegraphics{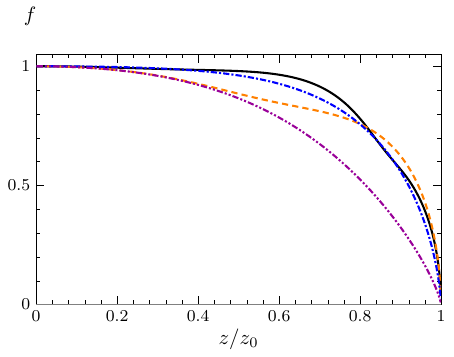}
    \caption{Examples of classical solutions in model A. The left (right) panel displays four numerical examples 
    of the background function $\phi$ ($f$), as a function of the holographic direction $z/z_0$, where $z_0$ is the end of space (dual to the IR regime of the field theory). We fix  \(\a=1\) and allow a few different choices of \(\f_c\). 
    We only show solutions with positive source \(\Lambda\) in the dual field theory, or equivalently with positive $\phi$ at small $z$; there are also solutions with negative \(\Lambda\) that can be obtained via the \(\mathbb{Z}_2\) transformation \(\f \to - \f\). For the two values of \(\f_c\) closest to the origin of the complex plane, we find ``bouncing'' solutions where the scalar field, \(\f\), changes sign (possibly multiple times). For larger values of \(|\f_c|\) the scalar field does not bounce, and has definite sign. }
    \label{fig:complex_fp_example_solutions}
    \end{center}
\end{figure}

We show in Fig.~\ref{fig:complex_fp_example_solutions} representative examples of the resulting solutions, keeping fixed  \(\a=1\), and varying the choice of \(\f_c\). We only show solutions with positive \(\Lambda\), for which $\phi$ is positive close to the UV boundary, $z\rightarrow 0$. For each value of \(\f_c\) there also exists an alternative solution with negative \(\Lambda\), related to the plotted ones by the \(\mathbb{Z}_2\) symmetry transformation \(\f \to - \f\). For \(\f_c = 1 + i\) (the dashed orange curve in the plot) we find a ``bouncing'' solution, in which \(\f\) changes sign at intermediate \(z\), and thus a solution which has \(\Lambda > 0\) with the IR behavior \(\f \to - \infty\). For \(\f_c = 0.5 + 0.5i \) (solid black) we also have a bouncing solution, this time with \(\f\) changing sign twice. On the other hand, for the other two values of \(\f_c\) plotted the field \(\f(z)\) grows monotonically. In general we find that the closer \(\f_c\) is to the origin of the complex plane, the more nodes there are where \(\f(z)\) changes sign.

\begin{figure}
        \includegraphics{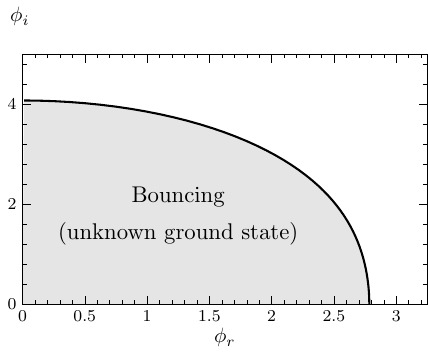}
    \caption{In model A, for \(\a=1\) and values of \(\f_c = \f_r + i \f_i\) inside the shaded shaded region, we find only bouncing solutions, in which \(\f(z)\) changes sign  at least once at an intermediate value of \(z\). In Sec.~\ref{sec:complex_fp_spectrum}, we show  that bouncing solutions exhibit a tachyonic mode in the spectrum of their fluctuations, and are thus unstable. We therefore do not have any candidate ground state in the shaded gray region. Along the boundary of the unstable region, indicated by the black curve, solutions with the IR asymptotics~\eqref{eq:complex_fp_ir_asymptotics_alt} have vanishing source, \(\Lambda=0\) in Eq.~\eqref{eq:complex_fp_background_uv}.}
    \label{fig:complex_fp_ground_states}
\end{figure}

In Fig.~\ref{fig:complex_fp_ground_states} we show, for \(\a=1\), the range of \(\f_c\) for which we find bouncing solutions, defined by the fact that $\phi$ changes sign at least once, contrasted to the region of parameter space for which the solutions have $\phi$ of fixed sign. We have not been able to find any other IR boundary conditions, and hence any other solutions, consistent with Eqs.~\eqref{eq:complex_fp_nested_eom}. Yet, we anticipate here a result presented  in the next subsection: the computation of the mass spectrum of fluctuations indicates that bouncing solutions exhibit a tachyonic mode, signaling their instability. They therefore cannot describe the ground state of the dual QFT, but as we have not been able to identify viable alternative solutions, we do not have a candidate ground state in the shaded region of parameter space in Fig.~\ref{fig:complex_fp_ground_states}. It is possible  that the ground state is not captured by our Poincar\'e-invariant ansatz in Eqs.~\eqref{eq:complex_fp_background_ansatz}, for instance it may be a state with spontaneously-broken translational invariance. Alternatively, there could be some intrinsic pathology in the field theory interpretation of the backgrounds in this region of parameter space. We leave this question open for future investigations, while focusing in the reminder of this section on values of \(\f_c\) leading to solutions for $\phi$ with a fixed sign. 

\begin{figure}
    \begin{center}
    \includegraphics{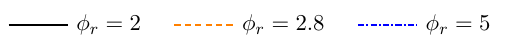}\\
    \begin{tabular}{r r}
    \includegraphics{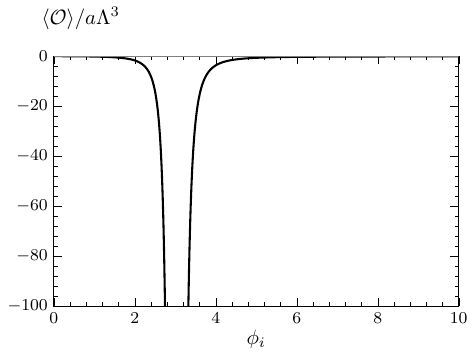} &
    \includegraphics{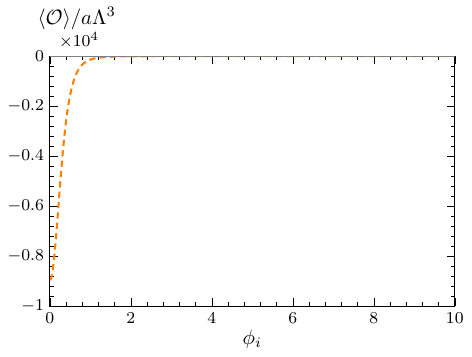}\\
    \includegraphics{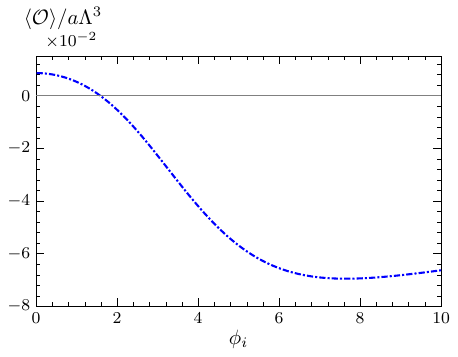}&
    \includegraphics{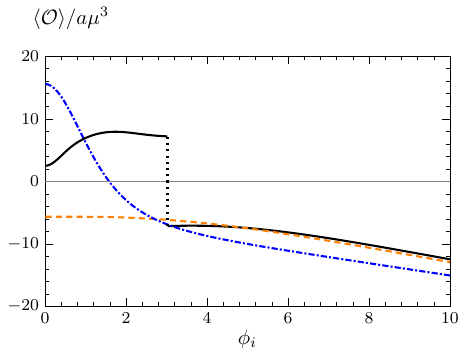}
    \end{tabular}
    \end{center}
    \caption{Numerical results for the expectation value \(\langle \cO \rangle\) of the scalar operator dual to the scalar field  \(\f\) in model A as a function of \(\f_i \), for several illustrative choices of \(\f_r \). The first three plots (reading left to right, top to bottom) show \(\langle \cO \rangle\) in units of the source, \(\Lambda\), while the bottom-right plot shows \(\langle \cO \rangle\)  in units of the scale, \(\m\), defined by the geometry. The top-left plot shows how, for \(\f_r=2\), we find that \(\langle \cO \rangle/\Lambda^3\) diverges at  \(\f_i \approx 3.025\). This is an artefact due to the vanishing of \(\Lambda\) for solutions with these choices of $\phi_c=\phi_r+i \phi_i$; for \(\f_r=3.025\) and \(\f_i \lesssim 3.025\) we find only bouncing solutions. The bottom-right panel demonstrates that there is a discontinuity in the response function, $\langle {\cal O} \rangle/(a \mu^3)$, when crossing the boundary displayed in Fig.~\ref{fig:complex_fp_ground_states}. }
    \label{fig:complex_fp_vev}
\end{figure}

In Fig.~\ref{fig:complex_fp_vev}, we show the expectation value, \(\langle \cO \rangle\), of the scalar operator dual to the field \(\f\) as a function of \(\f_i \), for \(\a=1\) and several representative choices of \(\f_r \). We show  \(\langle \cO \rangle\) in units of \(\Lambda\) as well as in units of $\mu$. We notice that \(\langle \cO \rangle/\Lambda^3\) diverges at \(\f_i \approx 3.025\) for $\f_r=2$. This is a point on the boundary displayed in Fig.~\ref{fig:complex_fp_ground_states}. The divergence is an artefact that occurs because  \(\Lambda\) vanishes precisely at this point of parameter space. The final plot in Fig.~\ref{fig:complex_fp_vev} shows \(\langle \cO \rangle\) in units of the intrinsic mass scale \(\m\) defined in Eq.~\eqref{eq:geometric_mass_scale}. It demonstrates the presence of  a discontinuity in the response function, $\langle {\cal O} \rangle/(a \mu^3)$, when crossing the boundary displayed in Fig.~\ref{fig:complex_fp_ground_states}. This occurs because, maintaining positive \(\Lambda\), the relevant solution switches between one with \(\f(z\to z_0) \to - \infty\) for \(\f_i \lesssim 3.025\) to one with \(\f(z\to z_0) \to + \infty\) for \(\f_i \gtrsim 3.025\). It would be tempting to identify the discontinuity as a signal of a phase transition. Unfortunately, as we cannot trust the bouncing solutions to be the correct ground states, we have to defer further to future studies on the presence of a possible phase transition and its physics implications.

\subsection{Spectrum of fluctuations}
\label{sec:complex_fp_spectrum}

We analyze the mass spectrum of perturbations of the metric and scalar field, by linearizing the equations of motion around   the classical solutions of model A described in the previous subsection. We define the small perturbations \(g_{mn} \to g_{mn} + \d g_{mn}\) and \(\f \to \f + \d \f\), and consider single Fourier modes with respect to the field theory directions \(x^\m\), with momentum \(p_\m\). For example, \(\d \f(z;x) = e^{i p_\m x^\m} \d \f(z;p)\). The spectrum of solutions for \(\d g_{mn}\) and \(\d \f\) that are normalizable at the asymptotically-\(AdS\) boundary and regular at the singularity at \(z = z_0\) lead to a discrete set of allowed values of \(m = \sqrt{- \h^{\m\n} p_\m p_\n}\), where \(\h^{\m\n}\) is the inverse of the Minkowski metric. These values of \(m\) correspond to the locations of poles in the momentum-space two-point functions involving the scalar operator, \(\cO\), and the stress tensor, \(T^{\m\n}\), in the dual field theory.

To determine the spectrum of allowed values of \(m^2\), we employ the gauge-invariant  formalism of Refs.~\cite{Bianchi:2003ug,Berg:2005pd,Berg:2006xy,Elander:2009bm,Elander:2010wd,Elander:2014ola}. This formalism---reviewed in appendix~\ref{app:fluctuations}---reduces the problem of determining the spectrum of metric and scalar fluctuations to that of finding UV-normalizable and IR-regular solutions to two decoupled equations, removing gauge artefacts. The spectrum of spin-zero excitations is determined by the solutions to the equations of motion for a gauge-invariant scalar variable, \(\mathfrak{a}\), formed from a linear combination of \(\d\f\) and one of the metric degrees of freedom, related to the trace of the fluctuations of the metric. As we show in appendix~\ref{app:fluctuations}, the equation of motion for \(\mathfrak{a}\) is
\begin{equation} \label{eq:complex_fp_spin0_eom}
    \frac{z^3}{ f^2} \le(\frac{f^2}{z^3}\mathfrak{a}'\ri)' 
    + \le(\frac{ m^2}{f} 
    - \frac{4 f^2 L^2 V_A \f'^2}{9 z^6 [(f/z^2)']^2}
    - \frac{4 L^2 f \p_\f V_A \, \f'}{3 z^3 (f/z^2)'}
    - \frac{L^2}{z^2} \p_\f^2 V_A \ri)\mathfrak{a} = 0.
\end{equation}
The spin-two states are  the transverse, traceless components of the metric perturbation, which we denote \(\mathfrak{e}_{\m\n}\), and obey the Klein--Gordon equation on the curved background metric in Eq.~\eqref{eq:complex_fp_background_ansatz}:
\begin{equation} \label{eq:complex_fp_spin2_eom}
   \frac{z^3}{f^2} \le( \frac{f^2}{z^3} \mathfrak{e}_{\m\n}'\ri)' +  \frac{m^2}{f} \mathfrak{e}_{\m\n} = 0.
\end{equation}

Near the boundary, for $z\rightarrow 0$,  solutions to the equations of motion~\eqref{eq:complex_fp_spin0_eom} and~\eqref{eq:complex_fp_spin2_eom} take the form
\begin{align}
    \mathfrak{a}(z) &= \mathfrak{a}^{(1)} z - \mathfrak{a}^{(1)} \le[\frac{m^2}{2} + \frac{\Lambda^2}{4} \le(3\a^2 - 2 - \frac{18}{\f_c^2} - \frac{18}{{\f_c^*}^2}\ri) \ri] z^3 \log(|\Lambda| z) + \mathfrak{a}^{(3)} z^3 + O(z^4 \log z),
    \nonumber \\
    \mathfrak{e}_{\m\n}(z) &= \mathfrak{e}^{(0)}_{\m\n} \le[1 + \frac{m^2}{4} z^2 + \frac{m^2}{16}\le(\frac{\Lambda^2}{3} - m^2\ri) \log(|\Lambda|z) \ri] + \mathfrak{e}_{\m\n}^{(4)} + O(z^6 \log z),
\end{align}
where \((\mathfrak{a}^{(1)}, \mathfrak{a}^{(3)}, \mathfrak{e}^{(0)}_{\m\n}, \mathfrak{e}^{(4)}_{\m\n})\) are integration constants fixed by the boundary conditions. Near the singularity at \(z=z_0\), the regular solution to the spin-zero equation of motion behaves as\footnote{For \(\a > \sqrt{2/3}\) the  IR expansion of \(\mathfrak{a}\) also includes a term proportional to \(m^2 s^{-16/3\a^2} e^{-2(1-2/3\a^2)s }/ (\a^2 - 2/3)\), similar to the expansion of \(\mathfrak{e}_{\m\n}\) in equation~\eqref{eq:complex_fp_fluctuation_ir_asymptotics_spin_two}, but this is always subleading compared to those in Eq.~\eqref{eq:complex_fp_fluctuation_ir_asymptotics_spin_zero}.}
\begin{equation} \label{eq:complex_fp_fluctuation_ir_asymptotics_spin_zero}
    \mathfrak{a}(z) = \mathfrak{a}^{(\mathrm{IR})} \le[1 + \frac{2}{s} + O(s^{-2}\log s) \ri],
\end{equation}
while the regular solution to the spin-two equation of motion behaves as
\begin{equation} \label{eq:complex_fp_fluctuation_ir_asymptotics_spin_two}
    \mathfrak{e}_{\m\n}(z) = \begin{cases}
        \mathfrak{e}_{\m\n}^{(\mathrm{IR})} \le[1 - \dfrac{m^2 z_0^2}{21 f_\mathrm{IR} s^7} + O(s^{-8} \log s) \ri],
        & {\rm if}~~ \a = \sqrt{2/3},
        \\[1em]
        \mathfrak{e}_{\m\n}^{(\mathrm{IR})} - \mathfrak{e}_{\m\n}^{(\mathrm{IR})} \dfrac{\a^4 z_0^2 m^2 s^{-16/3\a^2} e^{-2(1-2/3\a^2)s}}{2(\a^2 - 2/3)(\a^2 +4/3) f_\mathrm{IR}}  \le[ 1 + O(s^{-1} \log s) \ri],
        &{\rm if}~~  \a > \sqrt{2/3},
    \end{cases}
\end{equation}
where \(s=-\log(1-z/z_0)\), and \(\mathfrak{a}^{(\mathrm{IR})}\) and \(\mathfrak{e}_{\m\n}^{(\mathrm{IR})}\) are integration constants.
For given values of \(\a\) and \(\f_c\) (and of $\Lambda$), the mass spectrum of spin-zero fluctuations  is the set of values of \(m^2\) for which solutions to Eq.~\eqref{eq:complex_fp_spin0_eom} that are regular at \(z=z_0\) are also normalizable at the boundary, that is, have \(\mathfrak{a}^{(1)}=0\). Similarly, the mass spectrum of spin-two fluctuations is the set of values of \(m^2\) for which solutions to Eq.~\eqref{eq:complex_fp_spin2_eom} that are regular at \(z=z_0\) are also normalizable at the boundary, that is, \(\mathfrak{e}_{\m\n}^{(0)}=0\).

\begin{figure}[t]
    \begin{tabular}{c c}
         \includegraphics{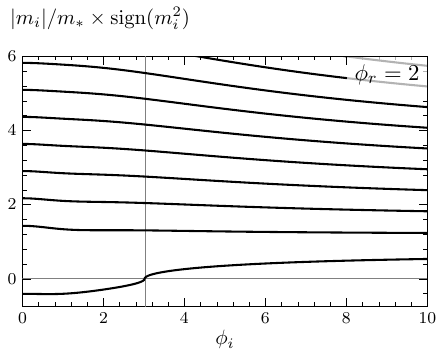}   &
         \includegraphics{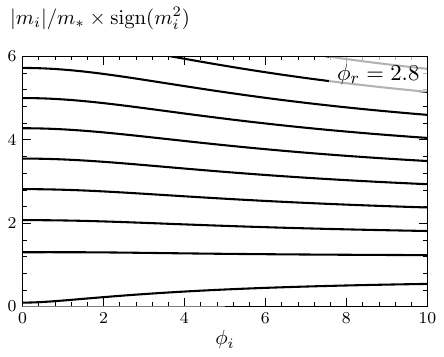}
         \\
         \includegraphics{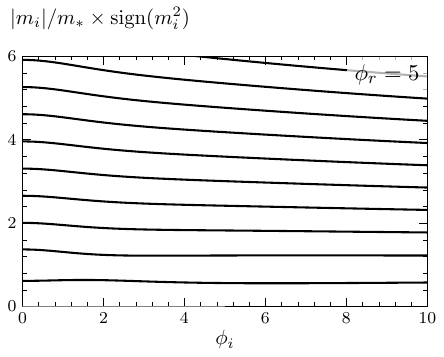} & 
         \includegraphics{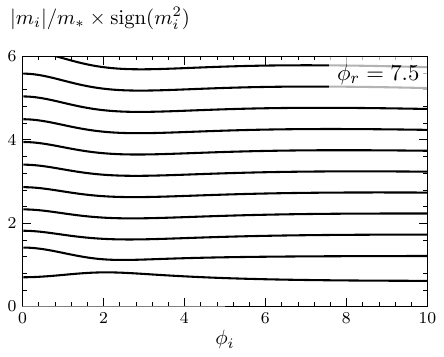} 
    \end{tabular}
    \caption{Mass spectra of spin-zero states, in model A, for \(\a=1\) and representative values of \(\f_r\). The masses are presented in units of the mass, \(m_*\), of the lightest spin-two mode. We plot the absolute values of the masses multiplied by the sign of their square, such that  negative values on the plot correspond to tachyonic modes with \(m^2 < 0\). In the top-left panel, which shows the spectrum for \(\f_r = 2\), the vertical gray line is at \(\f_i = 3.025\), and denotes a point of parameter space along the boundary identified in Fig.~\ref{fig:complex_fp_ground_states}. For \(\f_i\leq 3.025\) we find only bouncing solutions, which as shown by the plot exhibit a tachyonic mode. The other panels show choices of parameters for which monotonic solutions exist, and no tachyons emerge. We highlight the presence of a region of $\phi_i$, in the two top panels, for which the lightest scalar state has positive mass squared, yet it is anomalously light in comparison to the rest of the spectrum.}
    \label{fig:complex_fp_alpha1_spectrum}
\end{figure}

We compute the fluctuation spectra using a pseudospectral method, by adopting the following numerical strategy.  
We map the interval \(z\in [0,z_0]\) into the interval \(x \in [-1,1]\), and rescale the gauge-invariant fluctuations, by defining 
\begin{equation}
    x \equiv \frac{ \log(1-z/z_0) + 1}{ \log(1-z/z_0)-1}\,,
    \qquad
    \bar{\mathfrak{a}} \equiv \frac{1-x}{1+x} \mathfrak{a}\,,
    \qquad
    \bar{\mathfrak{e}}_{\m\n} \equiv (1-x) \mathfrak{e}_{\m\n}\,.
\end{equation}
We note that  the logarithmic behavior of the change of variable near \(z=z_0\) has been chosen in order to tame some of the non-analyticities appearing in the IR expansions of the fluctuations, so that solutions to the fluctuation equations may be better approximated as sums of Chebyshev polynomials in \(x\). In terms of these new variables, the boundary conditions can be equivalently rephrased as
\begin{equation} \label{eq:complex_fp_spectral_bcs}
    \bar{\mathfrak{a}}(x= \pm1) = \bar{\mathfrak{e}}_{\m\n}(x= \pm1) = 0.
\end{equation}
We then decompose \(\bar{\mathfrak{a}}\) and \(\bar{\mathfrak{e}}_{\m\n}\) as a sum of the first  \(N_\mathrm{grid}\) Chebyshev polynomials in \(x\), and demand that the fluctuation equations, \eqref{eq:spin0_eom} and~\eqref{eq:spin2_eom}, be satisfied on the grid of Chebyshev extrema \(x_k = \cos [k \pi / (N_\mathrm{grid} - 1)]\) for \(k = 1,2,\dots,N_\mathrm{grid}-2\), supplemented with the boundary conditions~\eqref{eq:complex_fp_spectral_bcs} at \(x=\pm 1\).  The numerical approximation to the spectrum of \(m^2\) is finally extracted in terms of the generalized eigenvalues of the resulting \(N_\mathrm{grid}\)-dimensional matrix equation for the Chebyshev expansion coefficients. Many of the resulting eigenvalues are spurious numerical artifacts. To eliminate the spurious eigenvalues and to check convergence, we compute the spectra for two different values of \(N_\mathrm{grid}\), and retain only modes that agree between the two computations. All spectra shown in this section have been computed for \(N_\mathrm{grid} = 70\) and \(N_\mathrm{grid}=80\). For further details of the pseudospectral method, see Refs.~\cite{boyd_book,Jansen:2017oag}.

\begin{figure}[t]
   \begin{tabular}{c c}
         \includegraphics{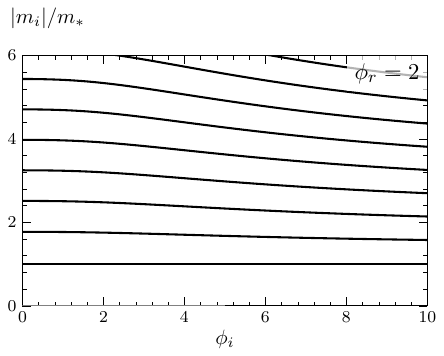}   &
         \includegraphics{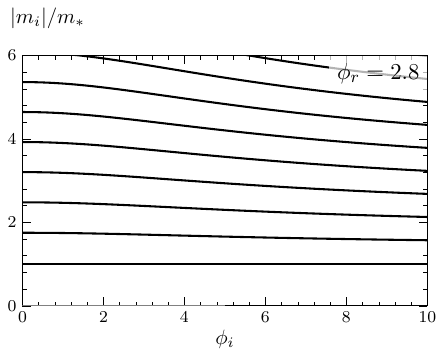}
         \\
         \includegraphics{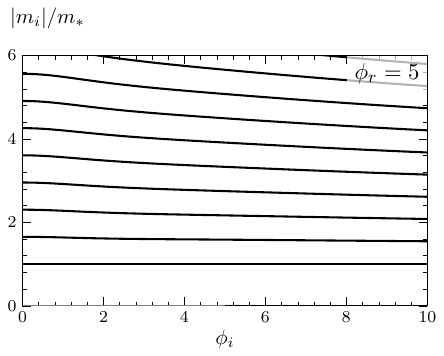} & 
         \includegraphics{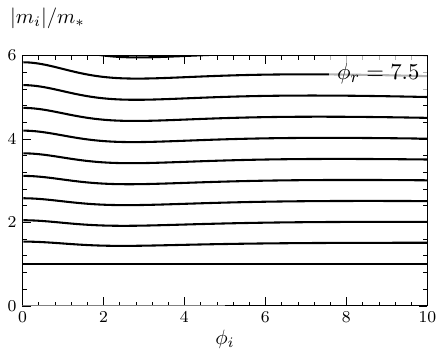}
    \end{tabular}
    \caption{Same as Fig.~\ref{fig:complex_fp_alpha1_spectrum}, but for spin-two states. The mass spectra of tensor modes do not show any striking features, but rather reproduce the expectations in the gapped field theory dual to a pure supergravity background, with masses of excited states growing linearly with the excitation number.
    \label{fig:complex_fp_alpha1_spectrum_spin2}}
\end{figure}

In Fig.~\ref{fig:complex_fp_alpha1_spectrum} we show the mass spectrum of spin-zero states computed for \(\a=1\) and few representative  values of \(\f_r\), as a function of \(\f_i\).  The spectra are shown in units of the lightest spin-two mass, which we denote by \(m_*\). We display them as their absolute value times the sign of \(m^2\). Hence, a mode that appears as negative in the plot corresponds to a negative value of of \(m^2\), corresponding to a tachyonic excitation in the dual QFT, while a mode that appears as positive corresponds to a positive value of \(m^2\).

A tachyonic exictation indicates that the gravitational solution and its dual QFT state are unstable. This is because every value of \(m^2\) in the spectrum corresponds to a family of excitations that behave with time as \(e^{\pm i \sqrt{m^2}\, \hat{p}_\m  x^\m}\) for both choices of the \(\pm\) sign, where \(\hat{p}\) is any unit-normalised momentum, \(\hat{p}^\m \hat{p}_\m = -1\). Thus when \(m^2 < 0\) one of the choices of the \(\pm\) sign in the exponent will correspond to a mode that grows exponentially with time, signalling an instability.

The top-left panel of Fig.~\ref{fig:complex_fp_alpha1_spectrum} shows the spectrum for \(\f_r = 2\). The vertical gray line in the plot denotes the choice \(\f_i = 3.025\), a point of parameter space along the boundary identified in Fig.~\ref{fig:complex_fp_ground_states}. For values of \(\f_i\) smaller than this we find bouncing solutions. As can be seen in this figure, to the left of the vertical line there is a tachyonic mode. This is an example of the anticipated phenomenon, demonstrating the instability of the bouncing solutions. As \(\f_i\) is increased through \(\f_i = 3.025\),  the lightest state  has positive mass squared. Just above \(\f_i = 3.025\) the mode is anomalously  light compared to all other modes in the spectrum. Note that precisely at the boundary of the bouncing region at \(\f_i = 3.025\) where the lightest spin-zero mode is massless, the scalar field solutions have \(\Lambda=0\) and non-zero \(\langle\cO\rangle\), indicating that scale invariance is spontaneously broken.

In the top-right panel of Fig.~\ref{fig:complex_fp_alpha1_spectrum}  we show the spectrum for \(\f_r = 2.8\). This value of \(\f_r\) is just large enough that there are no bouncing solutions for any value of \(\f_i\) (see Fig.~\ref{fig:complex_fp_ground_states}). We do not find a tachyon in the spectrum. For small \(\f_i\), the lightest spin-zero mode becomes parametrically light compared to any of the other spin-zero or spin-two modes. This is a consequence of the fact that \(\f_r = 2.8\) and small \(\f_i\) lies close to the range \(\f_c\) for which the solution bounces, where the lightest spin-zero mode would become tachyonic. By contrast, the plots in the bottom row of Fig.~\ref{fig:complex_fp_alpha1_spectrum} show the spin-zero spectra for \(\f_r =5\) and \(\f_r= 7.5\), in which cases there are no bouncing solutions. These spectra show comparatively little structure, matching the generic qualitative expectations of the spectrum in a gapped theory.

For completeness, in Fig.~\ref{fig:complex_fp_alpha1_spectrum_spin2} we show the spin-two spectrum for \(\a=1\) and the same values of \(\f_r\) as in Fig.~\ref{fig:complex_fp_alpha1_spectrum}. In contrast to the spin-zero case, the spin-two spectra do not exhibit significant qualitative differences for different values of \(\f_r\). We only notice that, for excited states, the masses (rather than their squares) grow linearly with the excitation numbers. This is to be expected in the supergravity approximation of the dual to a gapped field  theory.

All of the plots of spectra shown in this section were computed for \(\a=1\). We have computed spectra for other values of \(\a\) as well, but since they exhibit the same qualitative behavior we do not show them in the main text. We show plots of sample spin-zero spectra for \(\a=\sqrt{2/3}\) and \(\a=1.25\) in App.~\ref{sec:extra_model_A_spectra}.

\subsection{Model A in compendium}
\label{sec:mA}

To summarize our results, in this section we proposed a simple bottom-up holographic model, consisting of a real scalar coupled to gravity in five dimensions, built to reproduce main features of a confining (or at least gapped) dual field theory, which admits complex fixed points at which putative complex CFTs can be defined. We identified a region of values for  \(\f_c\) for which  the complex fixed points sit relatively close to the origin of the complex plane. In this case,  we find that all available background solutions for \(\f(z)\) are bouncing and unstable,  as the spectra of their fluctuations exhibit a tachyonic spin-zero mode. Moving \(\f_c\) away from the origin, the solutions eventually cease to be bouncing. At the same point, the mode that was tachyonic becomes stable. Close to the boundary between these two behaviors, the lightest scalar state is anomalously light relative to the other spin-zero and spin-two modes in the spectrum. Moving \(\f_c\) yet further away from the origin, the mode ceases to be light.
  
There are hence two main, encouraging messages this class of models seems to support. Firstly, it is possible to find a parametrically light scalar state in the spectrum of a gapped theory, which emerges in proximity to a boundary in parameter space at which a classical instability appears. Second, there is no immediate relation between the presence of complex fixed points near the real axis and such phenomena, but rather a more general type of limiting behavior appears to drive the dynamics.

It would be tempting to identify the classical instability with the presence of a phase transition, but in the context of these models we do not find any stable ground state, leaving open the task of performing a global stability analysis and a characterization of the phase structure. It would also be tempting to identify the light scalar, which emerges in the proximity of the boundary of parameter space we  discovered in Fig.~\ref{fig:complex_fp_ground_states}, with a light dilaton. However, the existence of a region of parameter space for which we find only unstable solutions suggests that we are missing some class of solutions to model A that would provide the ground state in the unstable region. It may be that a global stability analysis would then show this new class of solutions to be the true ground state also in the region where we currently find stable solutions exhibiting a light dilaton---this unfortunately happens to be the case in some top-down holographic models~\cite{Elander:2020ial,Elander:2020fmv,Elander:2021wkc}. For these reasons we will now turn to a different model, for which we have been able to obtain a seemingly more complete picture of the phase structure.

\section{Model B: smoothly shrinking circles and phase transitions}
\label{Sec:B}

The main difference between models of type A and B in this paper is that in the latter the space ends smoothly, and the geometry is regular for all values of the holographic direction, while in the former the backgrounds meet the modest requirement that the space ends in a ``good'' singularity, according to Gubser's criterion~\cite{Gubser:2000nd}. In model B, one of the gauge theory directions (both in the gravity picture and in its dual field theory interpretation) is a circle, which shrinks smoothly to zero size at the bottom of the geometry, as in Ref.~\cite{Witten:1998zw}. As anticipated in the introduction, we have multiple reasons to consider this scenario. First, it provides an instance of a regular confining geometry with a light scalar in the spectrum. Secondly, we want to demonstrate that the light scalar is present in a physically stable region of parameter space, even if it is nearby a weak phase transition. To do so requires a detailed analysis of the global stability properties of the ground state---its zero-temperature thermodynamics. Last but not least, we want to demonstrate that the light scalar is a dilaton.

To these purposes, after presenting the action of  model B and the solutions of interest, we compute the free energy and the spectrum of fluctuations. We also repeat the calculation of the spectrum of scalar fluctuations in the probe approximation, to show that  the results differ from those obtained in the fully back-reacted case, with the light scalar mode absent in the probe spectrum. This finding demonstrates that, in  regions of parameter space in proximity to the onset of a second-order phase transition, the lightest state in the theory has a composition that includes a mode coupled to the trace of the stress-energy momentum of the dual field theory: the dilaton.

\subsection{Models of type B}
\label{Sec:ModelB}

In models of type B, the dynamics of the gravitational theory is governed by a five-dimensional action that takes the form
\begin{equation}\label{eq:actionGH}
S\,=\,\frac{1}{16\pi G_5}\left\{\int_{\mathcal{M}} \dd^5x\sqrt{-g} \,\left[R-\frac12 g^{MN}\partial_M\phi\partial_N\phi
-V_B(\phi)\right]\, + \int_{\partial \mathcal{M}}\dd^4x \sqrt{-h} \, \left[2K+\frac{}{}4
\lambda_B(\phi)\right]\right\}\,,
\end{equation}
where $\phi$ is a real scalar, $V_B(\phi)$ its potential, and $R$ is the Ricci scalar associated to the metric $g_{MN}$, which has determinant $g$. Part of the action is localized at the boundary, $\partial \mathcal{M}$. It is expressed in terms of the induced metric, $h$, the extrinsic curvature, $K$, and the boundary-localized potential $\lambda_B(\phi)$. Throughout this section, capitalized  Latin indices \(M,N,\dots\) denote five-dimensional bulk directions.  The reason to do so is that we adopt an ansatz that assumes  one of the bulk directions to be a circle. We hence use lowercase indices \(m,n,\dots\) for the resulting, dimensionally reduced, four-dimensional bulk theory, and Greek indices \(\m,\n,\dots\) for its three-dimensional field-theory dual. From the action in Eq.~\eqref{eq:actionGH}, one derives the classical background equations,
\begin{equation}\label{eq:EOM}
    R_{MN} - \frac{1}{2}g_{MN} R = \frac{1}{2} \partial_M\phi \partial_N\phi -\frac{1}{2}g_{MN} \le[\frac{1}{2} (\p \f)^2 + V_B(\phi) \ri]\,,
    \qquad
    \nabla^2\phi \,=\,{\partial_ \phi  V_B(\phi)}\,,
\end{equation}
as well as boundary conditions, which we do not report explicitly for simplicity.

We  write  the potential  in terms of a superpotential, $W(\phi)$, which, with the normalization conventions of Eq.~(\ref{eq:actionGH}), obeys the relation
\begin{equation}
V_B(\phi)= -\frac{16}{3}W(\phi)^2 + 8 \left(\frac{\partial W(\phi)}{\partial \phi}\right)^2\,.
\end{equation}
We emphasize that this choice has nothing to do with supersymmetry, it just simplifies some of the equations, for example in the definition of boundary-localized potentials. We adopt the superpotential proposed in Ref.~\cite{Bea:2018whf}:
\begin{equation}\label{eq:super}
W(\phi) = \frac{1}{L}\left(-\frac{3}{2}-\frac{\phi^2}{8} -\frac{\phi^4}{64 \pM^2} + \frac{\phi^6}{64 \pQ}\,\right)\,,
\end{equation}
where $L$ has dimensions of a length and sets the curvature radius of the asymptotically-\(AdS\) solutions emerging in proximity of its stationary point $\phi=0$. The motivation for this choice of superpotential is that it is known that by tuning the parameters, $\pM$ and $\pQ$, one can explore a rich thermodynamic phase structure~\cite{Bea:2018whf}. Applications of this potential in the context of strongly coupled phase transitions include the physics of heavy-ion collisions~\cite{Bea:2021ieq}, cosmological phase transitions~\cite{Ares:2020lbt,Bea:2021zol,Bea:2021zsu,Bea:2022mfb}, and  primordial black hole production~\cite{Escriva:2022yaf}. For definiteness, we set  $\pQ= 10$ in the rest of our paper and let $\pM$ vary.

Expanding in powers of small \(\f\), the potential resulting from  Eq.~\eqref{eq:super} obeys
\begin{equation}
V_B(\phi) = \frac{1}{L^2}\parent{-12 -\frac{3}{2}\phi^2 -\frac{1}{12}\phi^4} + O(\phi^5).
\end{equation}
From the coefficient of the quadratic term, we read off that \(\f\) has mass-squared around the AdS solution $M_{\phi}^2L^{2}=-3  =-\Delta(4-\Delta) $. Applying the holographic dictionary,  the dual field-theory interpretation consists of a four dimensional conformal field theory deformed by an operator of dimension $\Delta=3$. 

In order to capture the features expected to emerge from a confining field theory within the gravity picture, we write the following ansatz for the background metric,
\begin{equation}\label{eq:Ansatz}
\dd s^2_{5} = \frac{L^2}{z^2} \le[
    \dd z^2 + f(z) \, \dd x_0^2 + g(z) \, (-\dd t^2 + \dd {x}_1^2 + \dd {x}_2^2) 
    \ri]\,,
\end{equation}
with $x_0\in [0,\per)$ a compact (periodic) direction. Our conventions are such that the boundary is at \(z=0\). In the solutions we consider, the internal circle parameterized by \(x_0\) shrinks smoothly to zero size at a particular value of the radial coordinate $z=\zs$.  This is reflected in the opening of a mass gap in the dual field theory, with a mass scale related to \(\zs\).

By substituting the ansatz in Eq.~\eqref{eq:Ansatz} into the equations of motion in Eqs.~\eqref{eq:EOM}, we find the following set of independent equations for the background fields, which are assumed to depend only on the holographic coordinate, $z$:
\begin{subequations} \label{eq:B_ansatz_eom}
    \begin{align}
        \frac{z^2 g''}{g} - \frac{3 z g'}{g} + \frac{z^2 \f'^2}{6} + \frac{L^2 V_B}{3} &= 0 \, ,
        \label{eq:B_ansatz_eom_g}
        \\
        z^2 \f'' - \le[1 - \frac{z g'}{g} - \frac{g (z^2 \f'^2 - 2 L^2 V_B)}{6 z^3 (g/z^2)'} \ri] z \f' - L^2 \p_\f V_B&= 0 
        \, ,
        \label{eq:B_ansatz_eom_phi}
        \\
        \frac{z f'}{f} + \frac{z g'}{g} - \frac{g (z^2 \f'^2 - 2 L^2 V_B)}{3 z^3 (g/z^2)'} - 4 &= 0\,.
        \label{eq:B_ansatz_eom_f}
    \end{align}
\end{subequations}
The prime denotes differentiation with respect to \(z\). The background function $f(z)$ does not appear in Eqs.~\eqref{eq:B_ansatz_eom_g} and~\eqref{eq:B_ansatz_eom_phi}, which determine \(g\) and \(\f\). This observation allows us to simplify the numerical procedure used to construct the solutions of interest. Equations~\eqref{eq:B_ansatz_eom} are invariant under the three independent rescalings \(f \to \Omega_f^2 \,f\), \(g \to \Omega_g^2 \,g\), and \(z \to \Omega_z \,z\), with \(\Omega_{f,g,z}\) constant parameters.

We demand that the geometry be regular for all values of \(z\). For $z$ close to \(\zs\), this is ensured if the function $f$  vanishes with the appropriate power of $(z-\zs)$, while $g$ and $\phi$  reach a finite value. At leading order in an expansion in small powers of  \((z-\zs)\), solutions to Eq.~\eqref{eq:B_ansatz_eom} behave as
\begin{equation} \label{eq:B_ir_asymptotics}
f (z)= \fs (z-\zs)^2+\cdots\,,\qquad g(z)= \gs +\cdots\,,\qquad \phi (z)= \psIR +\cdots\,,
\end{equation}
where \((\fs,\gs,\psIR)\) are integration constants and the ellipses denote terms that vanish faster when \(z \rightarrow \zs\). Regularity requires that the function $f(z)$ has a double zero at $z=\zs$. The coefficient $\fs$ of this zero is related to the period $\per$ of the compact coordinate $x_0$; to avoid a conical singularity at \(z =\zs\) we must impose \(\fs = (2\pi/\per)^2\).

We also restrict our attention to  solutions for which the local geometry is asymptotically AdS, by requiring that \(f\) and \(g\) tend to constant values and \(\f\) vanishes as \(z \to 0^+\). Using the scaling symmetries enjoyed by Eqs.~\eqref{eq:B_ansatz_eom}, these constant values can be fixed to \(f(0) = g(0) = 1\). Solutions to the equations of motion~\eqref{eq:B_ansatz_eom} supplemented by these boundary conditions obey the UV (small-\(z\)) expansions
\begin{eqnarray}
\label{UVexp_alt}
\begin{aligned}
\phi(z) &= \phi_s z + \phi_v z^3 + O(z^5)\,,\\
g(z) &= 1 - \frac{\f_s^2}{12} z^2 - \frac{1}{16} \le(q + 2 \f_s \f_v - \frac{\f_s^4}{18}\ri) z^4 + O(z^6)\,,\\
f(z) &= 1 - \frac{\f_s^2}{12} z^2 + \frac{1}{16} \le(3q - 2 \f_s \f_v + \frac{\f_s^4}{18}\ri) z^4 + O(z^6)\,,\\
\end{aligned}
\end{eqnarray}
where \((\f_s,\f_v,q)\) are integration constants and we have retained only terms up to the first appearance of these parameters in the expansion. The definition of \(q\) is chosen to simplify subsequent expressions. Through the holographic dictionary, we identify the leading integration constant, \(\f_s\), with the source for the scalar operator dual to \(\f\). The remaining integration constants, \(\f_v\) and \(q\), are determined by the requirement of regularity at \(\zs\). As we will see in the next section, \(\f_v\) determines the vacuum expectation value of the scalar operator.

The solutions of interest are known only numerically. We construct them by using the expansion~\eqref{eq:B_ir_asymptotics} in proximity of $z=\zs$, which corresponds to the deep IR regime of the dual field theory, to set up appropriate boundary conditions to solve the equations of motion. The solutions are extended towards smaller values of $z$, where they approach asymptotically the AdS geometry that captures the far UV regime of the dual field theory. For each value of the parameter \(\pM\) appearing in the potential, we can obtain different solutions. These are distinguished in the IR by their values of \(\psIR\) and in the UV by their values of the dimensionless combination \(\f_v / \f_s^3\).

Invariance of the equations of motion~\eqref{eq:B_ansatz_eom} upon the  rescaling \(g \to \Omega_g^2 \,g\) implies the existence of a radially conserved quantity, \(Q\), such that \(\p_z Q = 0\). Applying the Noether procedure, we find\footnote{In contrast, the conserved quantities corresponding to the invariance of the equations of motion under \(f \to \Omega_f^2 \,f\) and \(z \to \Omega_z \,z\) are identically zero.}
\begin{equation}\label{eq:B_radial_quantity}
    Q \equiv \frac{L^3}{z^3}\sqrt{\frac{g(z)}{f(z)}}  \le[g(z)f'(z) - f(z)g'(z)\ri]\,.
\end{equation}
The radial conservation law may be used to partially relate UV and IR quantities; evaluating \(Q\) using the UV expansions in Eq.~\eqref{UVexp_alt}, we find \(Q = L^3 q\), while evaluating \(Q\) using the IR asymptotics in Eqs.~\eqref{eq:B_ir_asymptotics} leads to an expression for \(Q\) in terms of \(\fs\) and \(\gs\), which we write as follows:
\begin{equation}
    q = - \frac{2 \sqrt{\fs \gs^3}}{\zs^3}\,.
\end{equation}

\subsection{Dimensional reduction, holographic renormalization and (zero-temperature) thermodynamics}
\label{Se:thermoB}

The background solutions we have constructed are independent of the compact coordinate, $x_0$. The direction described by $x_0$ shrinks at the end of space, where the dual field theory effectively becomes (2+1) dimensional (and confines). For this reason, it is convenient to perform the dimensional reduction of the theory. We write the five dimensional metric as 
\begin{equation}
\dd s^2_{5} = e^{-\chi/\sqrt{3}} \dd s_4^2 +e^{2\chi/\sqrt{3}} \dd x_0^2  \,,
\end{equation}
with the four-dimensional metric denoted $\dd s_4^2 = \mathsf{g}_{mn} \dd x^m\dd x^n$. Integrating over the compact direction while discarding a total derivative, the action in Eq.~\eqref{eq:actionGH} can be recast in the following form:
\begin{eqnarray}\label{eq:4D}
S &=& \frac{1}{16\pi G_4}\left[\int_{\mathsf{M}} \dd^4x\sqrt{-\mathsf{g}}\left(\mathsf{R}
-\frac12 \mathsf{g}^{mn}\partial_{m}\phi\,\partial_{n}\phi - \frac12\,
\mathsf{g}^{mn}\partial_{m}\chi\,\partial_{n}\chi-V(\phi,\chi)\right) \nonumber \right. \\
&&+  \left.\int_{\partial \mathsf{M}}\dd^3x \sqrt{-\mathsf{h}} \, 
\left(2\mathsf{K}\frac{}{}+4 e^{-{\chi}/{(2\sqrt{3})}}\lambda_B(\phi)\right)\right]\,,
\end{eqnarray}
with $V(\phi,\chi) = e^{-\chi/\sqrt{3}}V_B(\phi)$. All the geometric quantities, $\mathsf{g}$, $\mathsf{R}$, $\mathsf{h}$, and $\mathsf{K}$, correspond to the four-dimensional metric $\mathsf{g}_{mn}$. On the other hand, $G_4 = G_5/\per$, which becomes
\begin{equation}
G_4 = \frac{L^3}{2 \pi a \,\per}    
\end{equation}
when we translate to gauge theory quantities and relate the gravitational constant in five dimensions to the coefficient $a$ as in Eq.~\eqref{eq:complex_fp_Weyl_anomaly}.

We will adopt the following domain-wall ansatz for the four-dimensional metric 
\begin{equation}\label{eq:4Dmetric}
\dd s_4^2  = e^{2B}( - \dd t^2 + \dd {x}_1^2 + \dd {x}_2^2 )  + \dd \rho^2\,,
\end{equation}
the relation between the two sets of coordinates being
\begin{equation}
\dd \rho = -\left(\frac{L}{z}\right)^{\frac{3}{2}}{f(z)^{\frac{1}{4}}}\, \dd z\,,\qquad e^{2B} = \frac{L^3}{z^3} g(z)f(z)^{\frac{1}{2}}\,,\qquad e^{\chi/\sqrt{3}} = \frac{L}{z}f(z)^{\frac{1}{2}}\,.
\end{equation}

The background equations derived from Eq.~\eqref{eq:4D} coincide with Eqs.~\eqref{eq:EOM}, as long as the background fields do not depend on $x_0$. In the reduced theory, the geometry presents a curvature singularity at the end of space in the IR. In this sense it is similar to the solutions examined in Section~\ref{sec:fixed_point_annihilation}. However, in this case the uplift to five dimensions provides a resolution of the singularity into regular and smooth geometries.

The long-distance (thermodynamic) behavior of the theory is controlled  by the on-shell action, that is interpreted  in terms of the (thermodynamic) effective potential (or, equivalently,  free energy) of the dual field theory. The radial integration in $S$ extends from the bottom of the geometry, $z=\zs$,  to some finite value, $z = z_{\text{\tiny UV}}$, near the boundary at $z=0$.  Holographic renormalization is applied to  treat the UV divergences appearing in the $z_{\text{\tiny UV}}\rightarrow 0$ limit~\cite{Bianchi:2001kw, Skenderis:2002wp, Papadimitriou:2004ap}. The boundary-localized  contributions to the action, $S_{\text{\tiny ct}}$,  consist of counterterms that remove such divergences. Due to  the choice we made of describing the dynamics in terms of the superpotential in Eq.~\eqref{eq:super}, it is natural to adopt as renormalization prescription the choice $\lambda(\phi)=W(\phi)$. We can hence write the counterterms as
\begin{equation}
S_{\text{\tiny ct}} = \frac{1}{4\pi G_4} \int_{\partial \mathsf M} \dd^3x \sqrt{-\mathsf{h}}\, e^{-{\chi}/{(2\sqrt{3})}}W(\phi)\,.
\end{equation}
The renormalized, on-shell action, $S_{\text{\tiny ren}}$, is defined by taking the limit  $z_{\text{\tiny UV}}\rightarrow 0$: 
\beqs
S_{\text{\tiny ren}}&=&\lim_{z_{\text{\tiny UV}}\to 0} S_{(\lambda=W)}\, \nonumber \\
\label{eq:Sren2p1}
&=& \frac{ \mathrm{vol}(\mathbb{R}^{2,1})}{16\pi G_4}\lim_{z_{\text{\tiny UV}}\to 0} \sqrt{-\mathsf{h}} \left(\frac{4}{3}\mathsf{K} +  4 e^{-{\chi}/{(2\sqrt{3})}} W \right)\Big|_{z_{\text{\tiny UV}}}\,,
\eeqs
where $V_{1,2}= \int\dd t \, \dd x_1 \, \dd x_2$ is the (infinite) volume of the spacetime in the reduced theory. Because there is a shrinking cycle, Eq.~\eqref{eq:Sren2p1} does not receive any contribution from the IR at $z = \zs$.

\begin{figure}
\includegraphics[width=.45\textwidth]{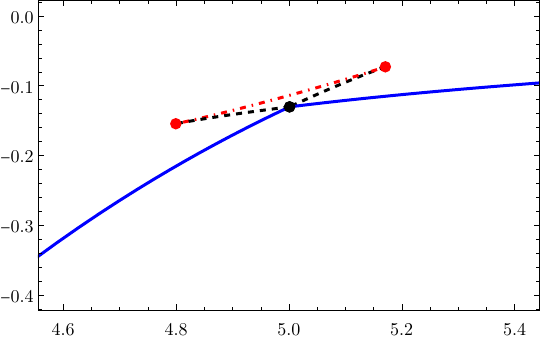} 
	\hspace{1cm}
	\includegraphics[width=.45\textwidth]{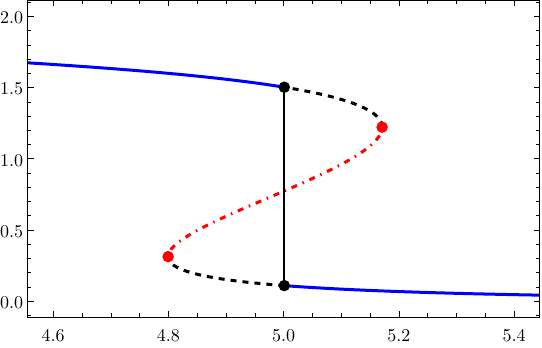}
	\put(-440,140){$\fQFT \times 8/(a\Lambda^3)$}
    \put(-210,140){$\langle\OO_\phi\rangle \times 8/(a\Lambda^2)$}
	\put(-350,-10){$\per\Lambda$}
	\put(-105,-10){$\per\Lambda$}
	\caption{\small  Free energy density (left) and vacuum expectation value of the operator dual to $\phi$ (right) for model B, with the choice $\pM= 97/100$, as a function of the deformation, $\per\Lambda$. In a range of parameters,  a first order phase transition appears, as indicated by the black dot in the classical swallow-tail diagrams.  Two stable branches of solutions meet and have the same free energy at the transition. At that point, the value of the response function,  $\langle\OO_\phi\rangle$, jumps discontinuously, as indicated by the vertical solid black line in the right plot. As illustrated in Fig.~\ref{fig:together}, one can smoothen out the transition by dialing the value of $\pM$.
\label{fig:thermobeta}}
\end{figure}

\begin{figure}
	\hspace{1cm}
	\includegraphics[width=.5\textwidth]{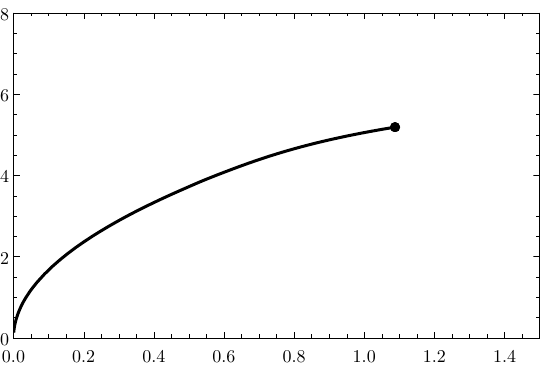}
	\put(-120,-10){$\pM$}
	\put(-240,165){$\per\Lambda$}
	\caption{\small The phase diagram of  model B. The solid curve indicates the critical value of $\per \Lambda$ as a function of $\pM$. The  disk denotes  the critical point, $(\pM^c,\per^c\Lambda) \simeq (1.088,5.186)$, at which the transition is   second order.}\label{fig:phase_space}
\end{figure}

According to the holographic dictionary, the free energy density of the dual theory is 
\begin{equation}
\fQFT = -\frac{S_{\text{\tiny ren}}}{ \mathrm{vol}(\mathbb{R}^{2,1})}\,,
\end{equation}
and the energy momentum tensor reads
\begin{equation}\label{eq:EandP4}
\mathsf{T}_{\ \nu}^{\mu} = -\frac{1}{{8\pi G_4}}\lim_{z_{\text{\tiny UV}}\to 0} \left[\sqrt{-\mathsf{h}} (\mathsf{K}_{\ \nu}^{\mu} -\delta_{\ \nu}^{\mu}(\mathsf{K} + 2e^{-{\chi}/{(2\sqrt{3})}}W))\Big|_{z_{\text{\tiny UV}}}\right] = \text{diag}(- \e, p, p)\,.
\end{equation}
Both can be expressed in terms of the coefficients appearing in the UV asymptotic expansions:
\begin{equation}
	p = - \fQFT = - \epsilon=  \frac{L^3}{16\pi G_4}\cdot \frac{1}{16} \left(-4 q - \frac{\ps^4}{\pM^2}+8\ps\pv\right)\,,\\
\end{equation}
while the vacuum expectation value of the operator dual to $\phi$ is
\begin{equation}
\begin{aligned}
\left\langle\OO_\phi\right\rangle &= \frac{1}{16\pi G_4}\lim_{z \to 0} z \sqrt{-\mathsf g} \left[ - \mathsf g^{zz}\phi'(z) - 4 e^{-{\chi}/{(2\sqrt{3})}}\partial_\phi W(\phi(z))\right] =  \frac{L^3}{16\pi G_4}\left(\frac{\ps^3}{4\pM} - 2\pv\right)\,.\\
\end{aligned}
\end{equation}
Exploiting these expressions we can characterize the different phases of the system. For each value of the parameter in the potential, $\pM$,\footnote{Recall that in this paper we fixed $\pQ = 10$ for convenience.} we can study the behavior of the free-energy density as a function of the source, $\Lambda=\phi_s$. We find that, for a range of choices of $\pM\in(0,\pM^c)$, there is evidence  of a  first-order quantum phase transition, as  exemplified in Fig.~\ref{fig:thermobeta}; the free-energy density is multivalued, as a function of the source $\Lambda$.  Remarkably, the line of first order phase transitions ends at $\pM^c$, as shown in Fig.~\ref{fig:phase_space}, with $\pM^c \simeq 1.088$.  A second order phase transition is present at this critical point. For values $\pM \gtrsim \pM^c$, the free energy becomes single valued and the theory just undergoes a smooth crossover. In the following subsection we investigate the spectrum near the second order phase transition. We will see that the end of the line of first order phase transitions has an important consequence: it triggers the appearance of a light dilaton in the vicinity of the critical point. 

\subsection{Spectrum of fluctuations}
\label{Sec:spectrumB}

As in Sec.~\ref{sec:complex_fp_spectrum}, we  study the mass spectrum of fluctuations of our solutions, using the gauge-invariant formalism of Refs.~\cite{Bianchi:2003ug,Berg:2005pd,Berg:2006xy,Elander:2009bm,Elander:2010wd}. It is convenient to employ this formalism in the four-dimensional bulk of the dimensionally reduced theory. We will restrict attention to fluctuations that are independent of the compact direction, neglecting the higher Kaluza--Klein (KK) modes. This should be sufficient for our goal of determining which solutions have a light scalar in their fluctuation spectra, since the excitations of the higher KK modes should have masses that are enhanced by a factor of order \(\ell^{-1}\), the inverse of the size of the compact direction, and therefore the spectra of excitations of the higher KK modes should not contain any modes lighter than the lightest excitations of the lowest KK mode.

The equations of motion for the gauge invariant fluctuations can be found in appendix~\ref{app:fluctuations}, for general bulk dimension and  an arbitrary number of scalar fields. To obtain the equations used in this section, one should set \(d=3\), $\Phi^1 =\chi$, and $\Phi^2 = \phi$ in the expressions in the appendix.

We compute the spectrum of spin-two and spin-zero states in dual field theory. We begin with the former. After removing spurious (gauge) degrees of freedom, one is left with a set of transverse, traceless components,  \(\mathfrak{e}_{\m\n}\), of the perturbation of the three-dimensional background metric in Eq.~\eqref{eq:4Dmetric}. These satisfy the Klein--Gordon equation on the background metric,
\begin{equation}\label{eq:spin2}
\left[\partial_{\rho}^2 + 3 (\partial_{\rho} B)\partial_{\rho} +m^2e^{-2 B} \right] \mathfrak{e}_{\mu\nu} (\rho)=0\,,
\end{equation}
where we have Fourier transformed with respect to the field theory directions, with corresponding momentum \(p_\m\), and with the definition $m^2 \equiv  -p^\mu p_\mu $. Each eigenvalue, \(m^2\), of this differential equation supplemented  by appropriate boundary conditions, is the mass squared of a spin-two state in the dual field theory.

It is convenient to analyze the boundary conditions by looking at the expansions of $ \mathfrak{e}_{\m\n}$  in the $z$ coordinate. In the deep IR, we consider solutions for which  $ \mathfrak{e}_{\m\n}$ is finite, which requires that its Taylor expansion reads
\begin{equation}\label{eq:spin2IR}
    \mathfrak{e}_{\mu\nu} = \spt \left[ 1 - \frac{m^2  }{4 \gs}(z-\zs)^2+\cdots\right]\,.
\end{equation}
In particular, each component of \(\mathfrak{e}_{\m\n}\) depends on one integration constant, ${\spt}$, the other integration constant having been eliminated by the requirement of regularity in the IR. 

In the far UV, two undetermined parameters, ${\sptUVo}$ and ${\sptUVf}$, appear in the asymptotic expansion of \(\mathfrak{e}_{\m\n}\):
\begin{equation} \label{eq:spin2UV}
    \mathfrak{e}_{\mu\nu} = \sptUVo + \frac{m^2\sptUVo}{4}z^ 2 + \sptUVf z^4 + \sptUVo \frac{ m^2 \left( \phi _s^2-3 m^2\right)}{48} z^4 \log z  +\cdots \,.
\end{equation}
The spectrum of spin-two modes is given by those values of $m^2$ for which a solution that is regular in the IR has $\sptUVo=0$, as these correspond to poles in the propagator of the traceless part of the stress tensor. 

The numerical strategy we follow to identify such values of $m^2$ is the following shooting method. We fix a value of $m^2$ and integrate the bulk equations along the holographic direction starting from the IR, using the expansion in Eq.~\eqref{eq:spin2IR} to set the initial conditions. Because the equations are linear, we can choose ${\spt} =1$ without loss of generality. Fitting the numerical solution of the differential equation to the UV expansion~\eqref{eq:spin2UV} we determine ${\sptUVo}$ for our choice of $m^2$. We repeat this procedure, scanning over different values of $m^2$. The mass spectrum of  spin-two states is given by the values of $m^2$ for which ${\sptUVo} = 0$. In the following, we denote the lightest spin-two state of a given theory by $m_*$. We use this value to set the scale in the mass spectrum of all other (spin-two and spin-zero) fluctuations. We do not show the full spectrum of spin-two states in model B, as it displays no distinguishing features---as in model A, see Fig.~\ref{fig:complex_fp_alpha1_spectrum_spin2}.

The spectrum of spin-zero fluctuations is found by studying two coupled second order differential equations for two scalars, $\mathfrak{a}^1$ and $\mathfrak{a}^2$, which are linear combinations of the perturbations of the scalar fields, $\phi$ and $\chi$, and scalar perturbations of the metric. As explained in appendix~\ref{app:fluctuations}, the equations of motion are given by
\begin{equation}\label{eq:spin0pertSimplified1}
       0=\left[\partial_{\rho}^2 + 3B' \partial_{\rho} + e^{-2B} m^2 \right] \mathfrak{a}^a -\left[\frac{V {\Phi^a}' \Phi_b'}{4B'^2}
    +\frac{V^a\Phi'_b+ {\Phi^a}' V_b}{2B'}+{V^a}_{b} \right]\mathfrak{a}^b\,,
\end{equation}
with \(\F^1 = \f\) and \(\F^2 = \chi\). Moreover, recall that \(V_a = \p V/\p \F^a\), \(V_{ab} = \p^2V/ \p \F^a \p \F^b\) and indexes are raised and lowered with the identity matrix, $\delta_{ab}$.

We solve Eqs.~\eqref{eq:spin0pertSimplified1} perturbatively in the deep IR, in terms of a Taylor expansion in powers of small $(z-\zs)$. Regularity near the end of space restricts the solutions to take the following form:
\begin{equation}\begin{aligned}\label{eq:spin0IR}
    \mathfrak{a}^1&= \spoIRi - \frac{1}{12}  \left(\frac{3 \spoIRi m^2}{\gs}-\frac{4
   \spoIRi L^2 \Vf(\pIR)}{\zs^2}-\frac{2\sqrt{3}\spoIRii L^2 \Vf'(\pIR)}{\zs^2}\right)(z-\zs)^2 + \cdots\,,\\
 \mathfrak{a}^2&= \spoIRii -\frac{1}{12}
    \left(\frac{3 \spoIRii m^2}{\gs}-\frac{2 \sqrt{3}  \spoIRi L^2 V_5'(\pIR)}{\zs^2} - \frac{3 \spoIRii L^2 V_5''(\pIR)}{\zs^2}\right)(z-\zs)^2 + \cdots\,,
\end{aligned}
\end{equation}
with two integration constants, $\spoIRi$ and $\spoIRii$, while $\gs$ is a background-dependent parameter.

We perform the same exercise in the far UV, near the boundary at $z\rightarrow 0$, by expanding in powers of small $z$. We find that the solutions take the form:
\begin{equation}\label{eq:ModelB_fluctuations_scalar_UV}
\begin{aligned}
 \mathfrak{a}^1&= \spooUVi + \frac{\spooUVi m^2 }{4}z^2+ \spofUVi z^4 +
    \frac{\spooUVi m^2 \left( \ps^2-3
   m^2\right)}{48 }z^4 \log z+\cdots \,,\\
  \mathfrak{a}^2&=\spooUVii z +2 \spotUVii z^3  - \frac{\spooUVii m^2}{2} z^3\log z+ \cdots\,,
\end{aligned}
\end{equation}
where the integration constants are denoted as $\spooUVi$, $\spooUVii$, $\spofUVi$, and $\spotUVii $.

\begin{figure}[t]\centering
	\includegraphics[width=.50\textwidth]{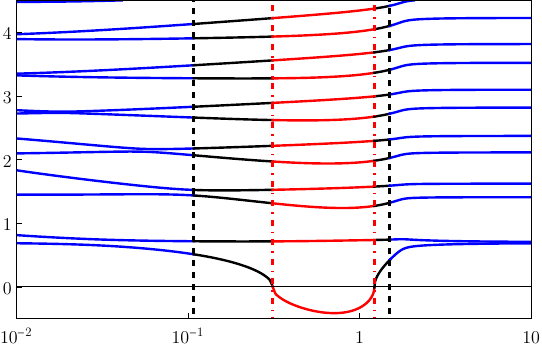}
	\put(-140,-10){\small$\langle\OO_\phi\rangle \times 8/(a\Lambda^2)$}
	\put(-260,165){\small$\displaystyle |m_i|\times\text{sign}(m_i^2)\,m_*^{-1}$}
	\caption{\small Mass spectrum of spin-zero states as a function of the vacuum expectation value of the operator dual to $\phi$, denoted  as $\langle {\cal O}_{\phi}\rangle$, normalized to powers of the scale $\Lambda$ and $a$, defined in Eq.~\eqref{eq:complex_fp_Weyl_anomaly}. The masses are normalized to the mass of the lightest spin-two state, $m_*$. The value of $\langle {\cal O}_{\phi}\rangle$ jumps at the first-order phase transition between the black, dashed vertical lines. Similarly, the red dot-dashed vertical lines indicate the value of  $\langle {\cal O}_{\phi}\rangle$ at the turning points, when plotted against $\per\Lambda$ (see Fig.~\ref{fig:thermobeta}). In this example, we set $\pM=0.97$.}\label{fig:perturbations}
\end{figure}

The mass spectrum of scalar fluctuations is obtained by identifying the discrete eigenvalues of $m^2$ for which there are solutions, of the form of Eq.~\eqref{eq:spin0IR}, which are regular in the IR and for which both $\mathfrak{a}_1(z)$ and $\frac{1}{z}\mathfrak{a}_2(z)$  vanish when $z\rightarrow 0$. Numerically, we adopt the method of Ref.~\cite{Kaminski:2009dh}. For each value for $m^2$, we construct two linearly independent numerical solutions in the bulk, which we denote $(\mathfrak{a}_+^1,\mathfrak{a}_+^2)$ and $(\mathfrak{a}_-^1,\mathfrak{a}_-^2)$. These are obtained by integrating Eq.~\eqref{eq:spin0pertSimplified1} starting  from the IR, with two different choices of the integration constants in the IR expansions~\eqref{eq:spin0IR}:
\begin{eqnarray}
    \mathfrak{a}^{1,2}_+: \quad (\spoIRi,\spoIRii) = (1,1)\,,
    \qquad \text{and} \qquad
    \mathfrak{a}^{1,2}_-: \quad (\spoIRi,\spoIRii) = (1, -1)\,.
\end{eqnarray}
The spin-zero spectrum corresponds to values of \(m^2\) for which there exists a linear combination of the plus and minus solutions for which both $\spooUVi =0$ and $\spooUVii = 0$ in Eq.~\eqref{eq:ModelB_fluctuations_scalar_UV}. This requirement if satisfied if the pairs $(\mathfrak{a}_+^1,z^{-1}\mathfrak{a}_+^2)$ and $(\mathfrak{a}_-^1,z^{-1}\mathfrak{a}_-^2)$ are proportional to each other, in the limit $z \to 0$. Thus, the spectrum consists of the values of $m^2$ for which the determinant of the matrix
\begin{equation}
    A(m^2)\equiv \lim_{z\to 0}\left(\begin{array}{cc}
          \mathfrak{a}^1_+(z)& \mathfrak{a}^1_-(z) \\
         z^{-1}\mathfrak{a}^2_+(z)& z^{-1}\mathfrak{a}^2_-(z) 
    \end{array}\right)
\end{equation}
vanishes.  In practice, we evaluate the determinant at a small, finite value of $z= z_\epsilon$, and then check convergence as $z_\epsilon$ is taken to zero.

Having performed the calculations of the spectra of fluctuations of both spin-zero and spin-two states, we can compare the results for different choices of background. For instance, in Fig.~\ref{fig:perturbations} we display the spectrum corresponding to the choice $\pM = 97/100$, the same value of \(\pM\) for which we showed the thermodynamics in Fig.~\ref{fig:thermobeta}. We find it convenient to normalize the mass of the fluctuations to that of the lightest spin-two state, $m_{\ast}$, and plot our results for different backgrounds by labeling them in terms of the value of $\langle {\cal O}_{\phi}\rangle$. A tachyonic state appears precisely at the turning points of Fig.~\ref{fig:thermobeta}, and persists over the whole range of (thermodynamically) unstable background solutions. We devote the next subsection to a detailed analysis of the relation between masses of the states, in particular that of the lightest scalars, and thermodynamics.

\begin{figure}[t]\centering
	\includegraphics[width=0.85\textwidth]{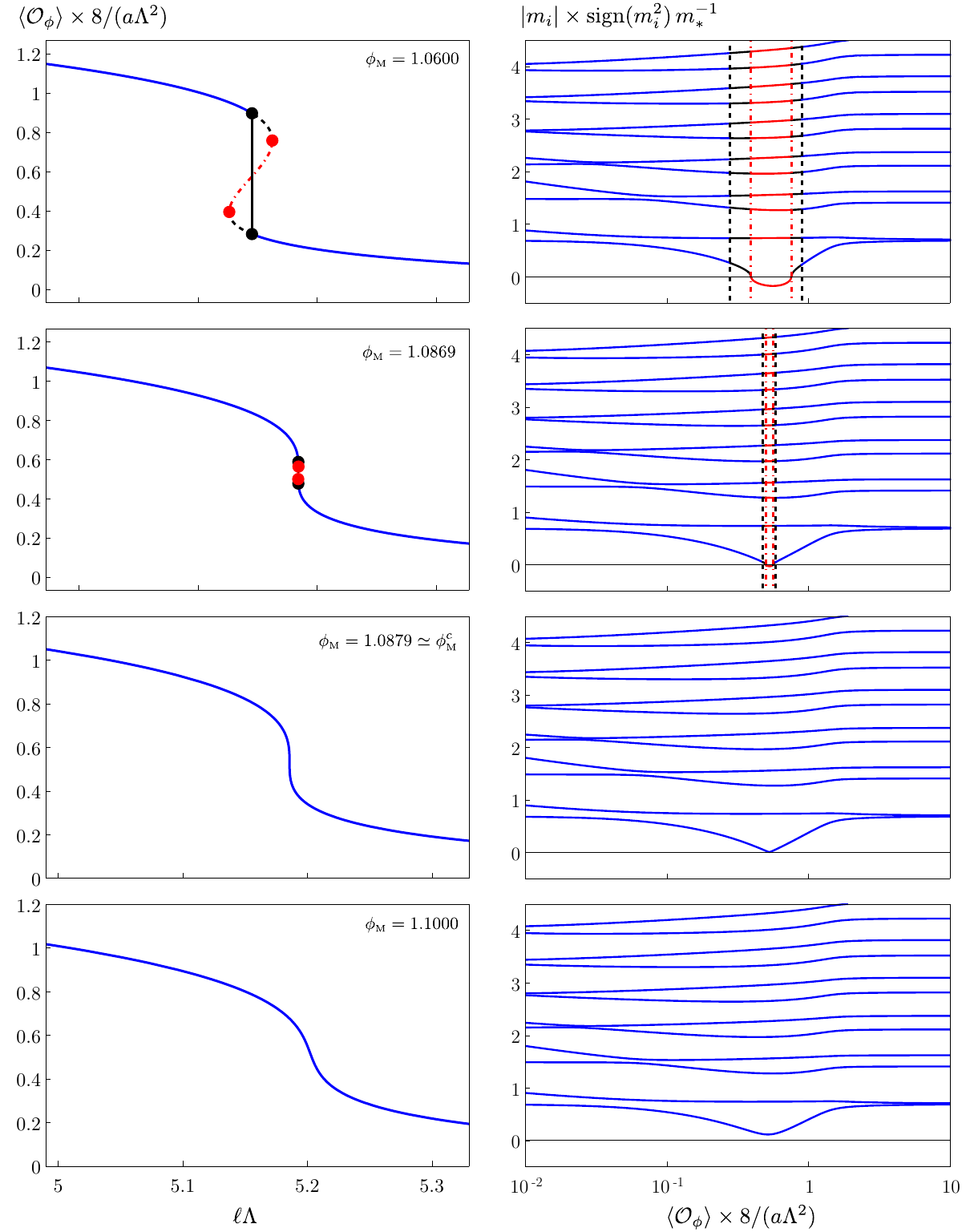}
	\caption{\small Vacuum expectation value of $\OO_\phi$ (left) and corresponding mass spectrum of fluctuations (right), for different choices of $\pM$ near the critical point. A massless scalar state appears at the critical point, $\pM=\pM^c$.
 }\label{fig:together}
 \end{figure}
 
\subsection{A light state near the critical point}
\label{Sec:lightB}

As we saw, having fixed the other parameters in the potential, generic choices of  $\pM < \pM^c$ lead to the appearance of  a first-order phase transition in the space of admissible, regular background solutions. For these, the free energy density plotted as a function of the source shows the typical multi-valued,  swallow-tail diagram (see Fig.~\ref{fig:thermobeta}) containing two branches of (meta-)stable solutions as well as one of unstable ones. In the spectrum, a tachyon appears along the unstable branch.

This subsection is devoted to discussing how such features change when one increases $\pM$ to approach and move beyond its critical value, $\pM^c$,  highlighted in Fig.~\ref{fig:phase_space}. Illustrative examples are depicted in Fig.~\ref{fig:together}, where the vacuum expectation value of the operator dual to $\phi$, $\langle\mathcal{O}_\phi\rangle$,  is displayed as a function of its source, $\Lambda$,  for several choices of $\pM$. The corresponding spectra are also shown. When the critical value $\pM^c$ is reached, the transition becomes of second order, as $\langle\mathcal{O}_\phi\rangle$ is continuous but its derivative, which diverges, is not. Ultimately, when the critical value is exceeded, a smooth crossover is encountered.

\begin{figure}[t]\centering
	\includegraphics[width=0.5\textwidth]{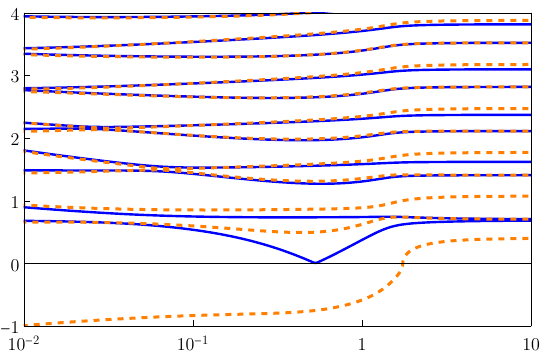}
    \put(-135,-10){\small$\langle\OO_\phi\rangle \times 8/(a \Lambda^2)$}
	\put(-260,160){\small$\displaystyle |m_i|\times\text{sign}(m_i^2)\,m_*^{-1}$}
	\caption{\small Spectrum of scalar fluctuations in model B computed in the probe approximation (orange dashed lines), defined by ignoring the metric fluctuations, compared  to the correct spectrum of scalar fluctuations (solid blue lines),	as a function of $\langle {\cal O}_{\phi}\rangle$. The calculation is performed by keeping fixed $\pM\simeq\pM^c$, as is the case for the third row of panels  in Fig.~\ref{fig:together}. When the fluctuations of the metric are ignored, the light state in the region near the critical point is missed by the probe calculation, which suggests it can be interpreted  as a light dilaton.}\label{fig:fig_probe_fluctuations}
 \end{figure}

As can be seen in the figure, for values of $\pM\lesssim\pM^c$ near the critical point but for which a phase transition is still present, the lightest state becomes parametrically light. This occurs along the physically realized and stable branch of solutions. Such a light state survives for $\pM\gtrsim\pM^c$, over a region of parameter space near the critical point. It hence appears that the presence of a nearby critical point in the space $(\pM,\per\Lambda)$ suppresses the mass of the lightest scalar state in the theory, compared to the other scales, and other masses.

We come  now to discussing the nature of such parametrically light scalar particle. If it is a dilaton, the pseudo-Nambu--Goldstone boson associated with the spontaneous breaking of approximate scale invariance, then it should couple to the trace of the stress-energy tensor---see for instance the lecture notes in Ref.~\cite{Coleman:1985rnk}. In the language of gauge/gravity duality, said trace sources the trace of the bulk fluctuations of the metric. But this is not a gauge-invariant statement, and furthermore the explicit breaking of scale invariance results in the mixing of the would-be dilaton with any other spin-zero, flavor-singlet excitation in the theory. Following the suggestion in Ref.~\cite{Elander:2020csd}, we redo the calculation of the spectrum of spin-zero states in the probe approximation. In doing so, we  ignore the back-reaction on the metric, and hence the metric fluctuations are ignored. This drastic approximation is doomed to miss the dilaton, while it should capture well the properties of states that are generic scalars. In the presence of modes that result from the admixture of the two, the approximation is going to work better for states that are predominantly the latter. The details of the computation are collected in Appendix~\ref{sec:app_probe_fluctuations}, and the results are shown in Fig.~\ref{fig:fig_probe_fluctuations}, for a particularly clear example, obtained with $\pM\simeq \pM^c$. In a range of values of $\langle {\cal O}_{\phi}\rangle$, the lightest scalar is approximately massless. The probe approximation reproduces all the excited states very well. However, it fails to capture the lightest mode, particularly when this state is very light. This result provides strong evidence that this light scalar state is to be interpreted as a light dilaton.

\subsection{Model B in compendium}
\label{sec:mB}

Model B is built starting from the same general set up as model A: a real scalar field is coupled to gravity in five dimensions. However, we adopt a simpler, polynomial potential, and
furthermore generalize the ansatz for the background, to admit solutions in which one of the dimensions is compactified on a circle. We then search for regular and smooth background solutions, in which the circle shrinks to zero size, hence modelling the generation of a mass gap in the dual, lower-dimensional field theory interpretation.

As in model A, we identified a boundary in the parameter space separating regions in which the background solutions are qualitatively different. By contrast with model A though, we were able to compute the free energy and perform a global stability analysis of the background regular solutions over the whole parameter space. This analysis demonstrated that the aforementioned boundary is actually a line of first-order, zero-temperature phase transitions, which terminates for a critical value of the tunable parameters in a second-order phase transition.

The calculation of the mass spectrum of fluctuations of model B shows that a tachyonic mode appears only in regions of parameter space leading to background solutions that are energetically disfavored, hence reinforcing the interpretation of the global stability analysis in terms of phase transitions in the theory. In the rest of the parameter space, both spin-zero and spin-two states display the properties one expects in a field theory that admits a supergravity dual description, with towers of approximately equally spaced masses.

The noticeable exception is represented by the region of parameter space close to the second-order phase transition: in this case, one of the spin-zero states is anomalously light, and its properties, analyzed by comparing the fully back-reacted calculation of the spectrum to the results of the probe approximation, are compatible with its interpretation as a dilaton, the parametrically light particle associated with spontaneous breaking of approximate scale invariance.

\section{Discussion}
\label{Sec:discussion}

Motivated by the long-standing problem of identifying classes of strongly coupled field theories with a spectrum of composite states containing a light dilaton---the pseudo-Nambu--Goldstone boson associated with spontaneous breaking of approximate scale invariance---in this paper we considered two classes of models, both constructed within the bottom-up approach to holography. We presented the mass spectrum of fluctuations and  the zero-temperature thermodynamic properties of the backgrounds we identified. These calculations provide two complementary (local and global) stability analyses of the vacuum structure of the dual field theories. We summarize in this final section our main findings, as well as their  implications for the more general research program on the dilaton. We also identify some future avenues for research.

In model A our main results are depicted in Figs.~\ref{fig:complex_fp_ground_states} and~\ref{fig:complex_fp_alpha1_spectrum}. We find evidence of the existence of a boundary, a hypersurface in the parameter space of the theory, that divides it into two regions within which the nature of the classical background solutions is qualitatively different.
Furthermore, we find evidence of the existence of an exactly massless scalar fluctuation along such boundary. In the proximity of such boundary, said state has an anomalously light mass compared to all other scales in the dual field theory. It is tempting to associate the boundary to the existence of a phase transition in the field theory at zero-temperature, and to associate the light state with a dilaton. Unfortunately, the nature of the model does not allow firm conclusions of this type. However, we do find a remarkable, robust result: while model A has been designed to exhibit the existence of complex fixed points, in the vicinity to which the RG flow slows down to the point that the physics is nearly conformal, their presence plays no central role in the aforementioned phenomena.
 
Our main results for models of type B are illustrated by Figs.~\ref{fig:thermobeta} to~\ref{fig:fig_probe_fluctuations}. Firstly, we find that there are regions of parameter space in which the lightest scalar in the spectrum of fluctuations has a mass parametrically suppressed with respect to all other mass scales in the theory. This happens even when the field-theory phenomena associated with confinement are modelled, in the gravity dual theory, by the smooth shrinking to zero size of an internal compact manifold (a circle), as suggested in the context of confining Yang--Mills theories~\cite{Witten:1998zw}. Furthermore, we find evidence that this scalar particle is approximately a dilaton. We think these are two important results, in the context of the historic discussion  in the literature~\cite{Holdom:1986ub,Holdom:1987yu,Appelquist:2010gy,Grinstein:2011dq}.
  
We find that this is not a generic expectation, though. As also discussed elsewhere in the literature, for instance in the context of top-down holographic models obtained within the context of extended supergravity theories in higher dimensions~\cite{Elander:2020ial,Elander:2020fmv,Elander:2021wkc}, the generic expectation is that when a gravity background is located in the proximity of a classical instability, then a  first-order phase transition in the parameter space of the field theory is also present. If so, generically one finds that along the physical, stable branches of admissible solutions, the lightest scalar state has mass that is not significantly lower than the scale of all other resonances. While it may be argued that the composition of such state does contain a significant dilaton component, yet its physical properties are not dissimilar to those of a generic scalar bound state.
 
The remarkable finding of this paper is that, as hinted to in the final paragraph of  Ref.~\cite{Elander:2021wkc} (see also references therein), we identified  a  model in which, for some choices of the parameters, the phase transition is weak; for such choices, we demonstrated the appearance of a light dilaton in the spectrum. Qualitatively, this expectation may be related to the ideas of Ref.~\cite{Kaplan:2009kr}. Nonetheless, the models discussed here include a mechanism for confinement, and the direct connection between the emergence of the dilaton and the proximity of a second-order phase transition appearing at the end of a line of first-order ones is novel in this context (a related argument can be found in the lattice literature~\cite{Lucini:2013wsa}). 
 
These findings further motivate the program of searching for realistic field theory realizations of the scenario in which a parametrically light dilaton appears in the spectrum of bound states of strongly coupled theories close to instabilities, in particular when critical points at the end of lines of first-order phase transitions exist. We do not have, yet, a systematic understanding of the conditions under which this might happen. Achieving this may require the study of broader classes of models than the two examples we provide in this paper. A prominent question is whether examples of this type exist in  the context of top-down holographic models, embedded in well understood supergravity theories that are believed to be the low-energy descriptions of fundamental theories of gravity (string theory).
 
In conclusion, the results of this paper demonstrate that the emergence of a light composite dilaton in strongly coupled gauge theories is possible in
the presence of confinement and phase transitions at zero temperature.The appearance of a dilaton in regions of parameter space near the end-point of lines of first order (zero-temperature) phase transitions is the central result that our examples illustrate. It would be interesting to demonstrate that this feature also survives in realistic field theories that can be studied in the context of lattice gauge theories, in which the discretized theories exhibit a rich, non-trivial phase structure~\cite{Lucini:2013wsa}.
Investigating these and closely related questions promises to be a fascinating future research program.

\section*{Acknowledgements}

We thank David Mateos for discussions and collaboration in the early stages of this project.

A.F. and C.H. are partially supported by the AEI and the MCIU through the Spanish grant PID2021-123021NB-I00 and by FICYT through the Asturian grant SV-PA-21-AYUD/2021/52177. The work of M.P. has been supported in part by the STFC Consolidated Grants No. ST/P00055X/1, No. ST/T000813/1, and ST/X000648. M.P. received funding from the European Research Council (ERC) under the European Union’s Horizon 2020 research and innovation program under Grant Agreement No.~813942.  The work of R.R. was supported by the European Union’s Horizon Europe research and innovation program under Marie Sklodowska-Curie Grant Agreement No.~101104286. Nordita is supported in part by Nordforsk.

\vspace{1.0cm}
{\bf Research Data Access Statement}---The data generated for this manuscript can be downloaded from  Ref.~\cite{data_release}.
\vspace{1.0cm}

{\bf Open Access Statement}---For the purpose of open access, the authors have applied a Creative Commons Attribution (CC BY) licence  to any Author Accepted Manuscript version arising.

\appendix

\section{More about model A}

In this appendix, we provide details on the procedure of  holographic renormalization applied to the models of type A we consider in the body of the paper.
For completeness, we also briefly expand the discussion of the  vacuum structure,  to consider cases other than $\alpha=1$, which are the focus of the main body of the paper.

\subsection{Holographic renormalization}
\label{app:complex_fp_holo_ren}

Using the equations of motion, the Lagrangian density appearing in the bulk part of the action, in Eq.~\eqref{eq:complex_fp_bulk_action}, reduces to a total derivative when evaluated on-shell:
\begin{equation}
    S_\mathrm{bulk}^\star = - \frac{L^3}{16 \pi \gn} \int \diff^4 x \, \int_\e^{z_0} \diff z \,  \le[\frac{f}{z} \le( \frac{f}{z^2} \ri)' \, \ri]'
    =- \frac{L^3}{16 \pi \gn} \int \diff^4 x \le[\frac{f}{z} \le(\frac{f}{z^2} \ri)' \ri]^{z=z_0}_{z=\e}\,,
\end{equation}
Here \(S_\mathrm{bulk} \equiv S - S_\mathrm{bdy}\), and \(\e\) is a small-\(z\) cutoff. Substituting the small-\(z\) expansion of \(f(z)\) from Eq.~\eqref{eq:complex_fp_background_uv_f}, we find that the regularized on-shell action is
\begin{equation}
    S_\mathrm{bulk}^\star = - \frac{L^3}{8 \pi \gn}  \int \diff^4 x \le[\frac{1}{\e^4} - \frac{\Lambda^2}{12 \e^2} + \frac{\Lambda^4}{32} \le(\frac{1}{3} - \frac{\a^2}{2} + \frac{3}{\f_c^2} + \frac{3}{{\f_c^*}^2}\ri)  \ri]+ O(\e).
\end{equation}

The boundary term in the action~\eqref{eq:complex_fp_bulk_action} receives two additive contribution, \(S_\mathrm{bdy} = S_\mathrm{GHY} + S_\mathrm{ct}\). The first is the Gibbons--Hawking--York (GHY) boundary term evaluated on the cutoff surface at \(z=\e\), \(S_\mathrm{GHY} = \frac{1}{8\pi \gn} \int \diff^4 x \, \sqrt{-h} \, K\), where \(h_{\m\n}\) is the induced metric on the cutoff surface, and \(K\) is its extrinsic curvature. Evaluating on-shell, we find
\begin{equation}
    S_\mathrm{GHY}^\star = \frac{L^3}{8 \pi \gn} \int \diff^4 x \le[
        \frac{4}{\e^4} - \frac{\Lambda^2}{3 \e^2} + \frac{ \Lambda^4}{8} \le(\frac{1}{3} - \frac{\a^2}{2} + \frac{3}{\f_c^2} + \frac{3}{{\f_c^*}^2}\ri)
        \ri] + O(\e).
\end{equation}
The second contribution to the boundary term, \(S_\mathrm{ct}\), is a set of local counterterms that cancel the divergences in the on-shell action,
\begin{equation} 
    S_\mathrm{ct} = - \frac{1}{8 \pi \gn L} \int \diff^4 x \, \sqrt{-h} \, \le[
        3 +\frac{\f^2}{4} + c \f^4
        + \frac{\f^4}{8}  \le(\frac{1}{3} - \frac{\a^2}{2} + \frac{3}{\f_c^2} + \frac{3}{{\f_c^*}^2}\ri) \log \f
    \ri],
    \label{eq:counterterms}
\end{equation}
where \(c\) is a free coefficient, since it multiplies a finite counterterm. Different choices of \(c\) amount to different renormalization schemes. The remaining coefficients in Eq.~\eqref{eq:counterterms} have been chosen to precisely cancel the small-\(\e\) divergences in the action; adding the counterterms, we find that the total renormalized on-shell action, \(S^\star = S_\mathrm{bulk}^\star + S_\mathrm{GHY}^\star + S_\mathrm{ct}^\star\), is given by
\begin{equation}
    S^\star  =  \frac{L^3}{32\pi\gn} \int \diff^4 x \le[\Lambda \F  - 4 c \Lambda^4 +\frac{ \Lambda^4}{8} \le(\frac{1}{3} - \frac{\a^2}{2} + \frac{3}{\f_c^2} + \frac{3}{{\f_c^*}^2}\ri)
    \ri].
\end{equation}

The one-point function of the operator, \(\cO\), dual to the bulk scalar field, \(\f\), is obtained from the functional derivative of the on-shell action with respect to \(\L\),
\begin{equation} \label{eq:scalar_vev_formula}
    \langle\cO\rangle = - \frac{\d S^\star}{\d \Lambda}.
\end{equation}
To obtain the contributions from the un-renormalized bulk action and the GHY term, we write the latter as a bulk integral over a total derivative and define~\(S_\mathrm{bulk} + S_\mathrm{GHY} = \int \diff^4 x \int \diff z \, \mathcal{L}\) for some Lagrangian \(\mathcal{L}\). Then, under a small variation \(\f \to \f + \d\f\) and \(g_{mn} \to g_{mn} + \d g_{mn}\) we find that the corresponding variation in the un-renormalized action is
\begin{align}
    \d S_\mathrm{bulk}^\star + \d S_\mathrm{GHY}^\star &= - \int \diff^4 x \, \le. \le( \d\f \frac{\p \mathcal{L}}{\p \f'}  + \d g_{mn} \frac{\p \mathcal{L}}{\p g_{mn}'}\ri)\ri|_{z=\e} 
    \nonumber \\
    &= \frac{L^3}{32 \pi \gn} \le[
        -\frac{2\Lambda\, \d \Lambda}{\e^2}
        +3 \F \, \d \Lambda - \Lambda \, \d \F
        +\frac{3\Lambda^3 \, \d \Lambda}{2}  \le(\frac{1}{3} - \frac{\a^2}{2} + \frac{3}{\f_c^2} + \frac{3}{{\f_c^*}^2}\ri)
    \ri] ,
\end{align}
where in the second equality we have substituted the UV expansions in Eqs.~\eqref{eq:complex_fp_background_uv}, and we have dropped terms that vanish in the limit \(\e \to 0\). The variation in the counterterms may be obtained straightforwardly by expanding Eq.~\eqref{eq:counterterms}:
\begin{equation}
    \d S_\mathrm{ct}^\star  = \frac{L^3}{32 \pi \gn} \int \diff^4 x \le[
        \frac{2\Lambda \, \d \Lambda}{\e^2}
        + \F \, \d \Lambda+ \Lambda \, \d \F - 16 c \Lambda^3 \, \d \Lambda
        - \Lambda^3 \, \d \Lambda  \le(\frac{1}{3} - \frac{\a^2}{2} + \frac{3}{\f_c^2} + \frac{3}{{\f_c^*}^2}\ri)
    \ri] .
\end{equation}
The divergent term and  the term proportional to \(\d\F\) cancel from the variation of the full renormalized action, as expected. Using Eq.~\eqref{eq:scalar_vev_formula} we find the one-point function:
\begin{equation}
    \langle\cO\rangle = - \frac{L^3}{8 \pi \gn} \le[\F  - 4 c \Lambda^3 + \frac{\Lambda^3}{8}
    \le(\frac{1}{3} - \frac{\a^2}{2} + \frac{3}{\f_c^2} + \frac{3}{{\f_c^*}^2}\ri)
    \ri].
\end{equation}
It is convenient to choose a renormalization scheme (a value of \(c\)) to cancel the terms in \(\langle\cO\rangle\) proportional to \(\Lambda^3\). In this scheme we find
\begin{equation}
    S^\star = \frac{L^3}{32 \pi \gn}  \Lambda \F \,  \mathrm{vol}(\mathbb{R}^{3,1}),
    \qquad
    \langle\cO\rangle = - \frac{L^3}{8 \pi \gn} \F,
\end{equation}
where \( \mathrm{vol}(\mathbb{R}^{3,1})\) denotes the regularized volume of Minkowski space, arising from the integral over the field theory directions in \(S^\star\).

\subsection{On the vacuum and spectra of model A}
\label{sec:extra_model_A_spectra}

\begin{figure}[htbp]
\begin{center}
\begin{tabular}{c c}
     \includegraphics{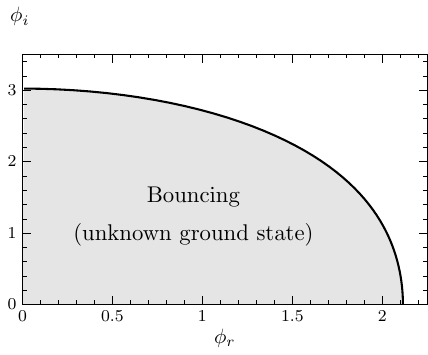}
     &
     \includegraphics{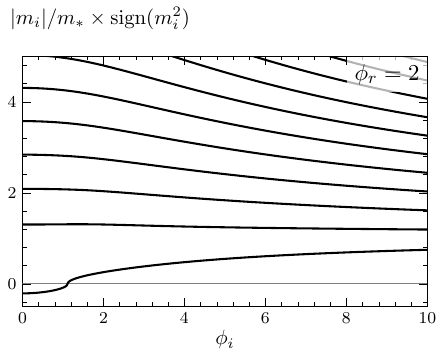}
     \\
     \includegraphics{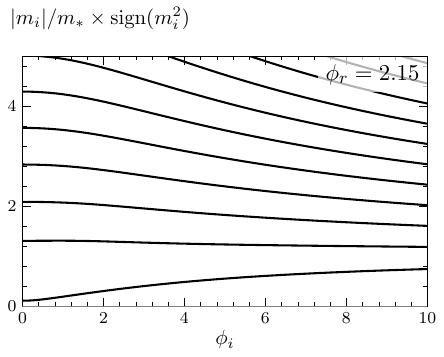}
     & 
     \includegraphics{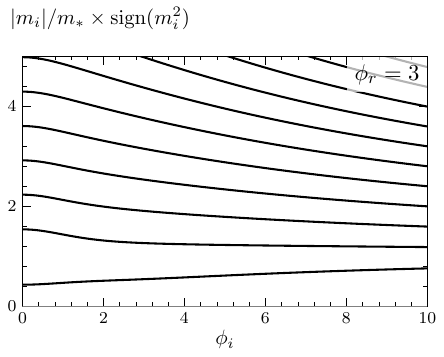}
\end{tabular}
    \caption{The phase space and spectra of model A for \(\a=\sqrt{2/3}\). \label{fig:sup1}}
    \end{center}
\end{figure}

\begin{figure}[htbp]
\begin{center}
\begin{tabular}{c c}
        \includegraphics{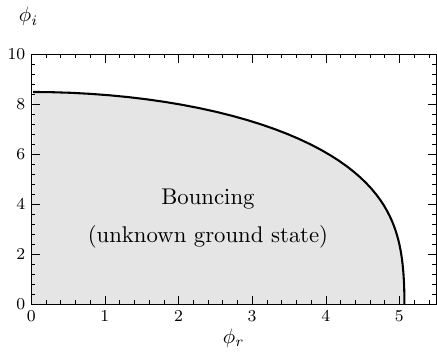}
        &
        \includegraphics{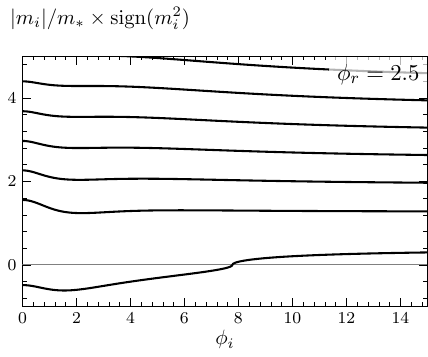}
        \\
        \includegraphics{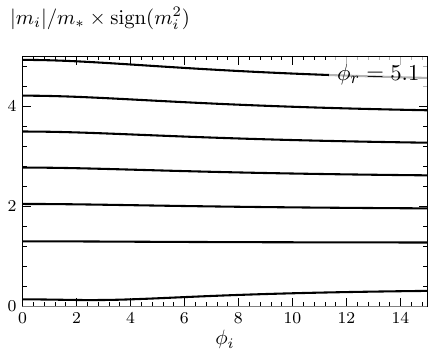}
        &
        \includegraphics{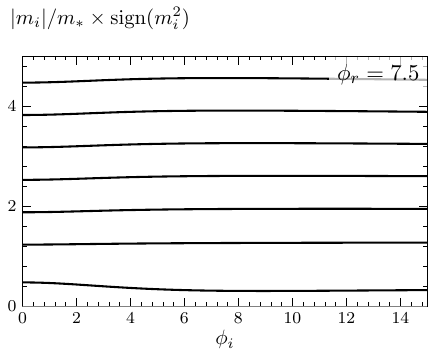}
\end{tabular}
\end{center}
    \caption{The phase space  and spectra of model A for \(\a=1.25\). \label{fig:sup2}}
\end{figure}

While in the main body of the paper we restricted attention  in model A to the choice $\alpha=1$,
we performed an extensive study of the parameter space to check that our findings have general value.
To illustrate this point, we present in Figs.~\ref{fig:sup1} and~\ref{fig:sup2} two numerical examples of the phase space of the models, with $\alpha=\sqrt{2/3}$ and $\alpha=1.25$, respectively, and samples of their corresponding spin-zero spectra. We plot the spectra as functions of \(\f_i\) for three different values of \(\f_r\): one value of \(\f_r\) for which our solutions are bouncing at small \(\f_i\), one value of \(\f_r\) for which our solutions are just outside the bouncing region, and one value of \(\f_r\) well away from the bouncing region.
As can be seen in the figures, by comparing them with Figs.~\ref{fig:complex_fp_ground_states} and~\ref{fig:complex_fp_alpha1_spectrum},
the qualitative features of our analysis are not affected by the change in $\alpha$. Namely, we find a region around the origin of the complex \(\f_c\) plane for which our solutions bounce and are unstable. Just outside this region the lightest spin-zero state is much lighter than the other states in the spectrum.

We also investigated the probe limit of spin-zero fluctuations in model A, similar to the analysis for model B performed in Sec.~\ref{Sec:lightB}. In this case the probe limit equation of motion is incompatible with imposing regularity of the spin-zero fluctuation in the IR---instead in the probe limit there are a pair of modes with complex exponents near \(z=z_0\)---so the spectrum is not well defined. One possible approach is to cut off the deep IR, considering only the region \(z \leq z_0 - \e\) with some small \(\e\). One can then compute the spectrum by imposing Dirichlet boundary conditions at \(z=z_0 + \e\), and investigate whether the spectrum converges in the limit \(\e\to0\). We find that this process indeed gives a convergent result for model A, however the result bears no relation to the true spectra plotted in Fig.~\ref{fig:complex_fp_alpha1_spectrum}.

\section{Fluctuation equations}
\label{app:fluctuations}

The spectrum of bound states is computed by studying the fluctuations 
of the background fields~\cite{Witten:1998zw} in a sigma-model coupled to gravity (see, for instance, Refs.~\cite{Brower:2000rp,Wen:2004qh,Kuperstein:2004yf}),
and makes use of a formalism that exploits gauge invariance~\cite{Bianchi:2003ug,Berg:2005pd,
Berg:2006xy,Elander:2009bm,Elander:2010wd,Elander:2014ola}. 
In this appendix we discuss the derivation of the equations of motion for the gauge invariant combinations of metric and scalar perturbations appearing in the main text. We follow Ref.~\cite{Berg:2005pd}.  We consider gravity in \((d+1)\) dimensions, minimally coupled to some number of scalar fields \(\F^a\), with bulk action
\begin{equation} \label{eq:fluctuation_equations_action}
    S_\mathrm{bulk} = \frac{1}{16 \pi G_{d+1}} \int \diff^{d+1} x \, \sqrt{-g} \le[R - \frac{1}{2} g^{mn} \p_m \F_a \p_n \F^a - V(\F)\ri],
\end{equation}
where \(V(\F)\) is the potential for the scalar fields, including a negative cosmological constant. Note our sigma-model metric is the identity, $\F_a = \delta_{ab} \F^b$. We  consider linearized fluctuations about an asymptotically AdS\(_{d+1}\) background geometry of the form
\begin{equation} \label{eq:fluctuation_equation_background}
    \diff s^2 =  \diff \r^2 + e^{2B(\r)} \h_{\m\n} \diff x^\m \, \diff x^\n,
    \qquad
    \F^a = \F^a(\r),
\end{equation}
where \(\r\) is the radial direction and \(x^\m\) are the field theory directions. Greek indices in this appendix are raised and lowered with the Minkowski metric, \(\h_{\m\n}\), and its inverse. The notation used in this appendix differs slightly from that used in section~\ref{sec:fixed_point_annihilation}; at the end of the appendix we provide a dictionary for translating between the two.

When perturbing the background in Eqs.~\eqref{eq:fluctuation_equation_background}, it is convenient to decompose a general metric in radial slicing according to the ADM formalism:
\begin{equation}
    \diff s^2 = \le(n^2 + n_\m n^\m\ri) \diff \r^2 + 2 n_\m \diff \r \, \diff x^\m + \bar{g}_{\m \n}  \diff x^\m \, \diff x^\n,
\end{equation}
where \((n,n_\m, \bar{g}_{\m\n})\) depend on \(\r\) and \(x^\m\). We rewrite the small perturbations  as
\begin{align}
    \bar{g}_{\m\n} &\to e^{2B(\r)}\le(\h_{\m\n} + h_{\m\n}\ri),
    &
    n_\m &\to \n_\m,
    \nonumber \\
    n &\to 1 + \n,
    &
    \F^a &\to \F^a(\r) + \y^a\,.
\end{align}
The small \((h_{\m\n},\n_\m,\n,\y_a)\)  depend on both the radial coordinate \(\r\) and the field theory coordinates \(x^\m\). We further decompose the symmetric tensor, \(h_{\m\n}\), as
\begin{equation}
    h_{\m\n} = t_{\m\n} + \p_\m u_\n + \p_\n u_\m + \frac{\p_\m \p_\n}{\Box} H + \h_{\m\n} h,
\end{equation}
where \(t_{\m\n}\) is a symmetric, transverse (\(\p^\m t_{\m\n} = 0\)), traceless tensor, \(u^\m\) is  transverse vector, and \(H\) and \(h\) are scalars. The field theory d'Alembertian \(\Box\equiv \h^{\m\n} \p_\m \p_\n\) is constructed using the Minkowski metric. If we define the transverse projector
\begin{equation}
    \Pi_{\m\n} = \h_{\m\n} - \frac{\p_\m \p_\n}{\Box},
\end{equation}
then we can extract the individual components of \(h_{\m\n}\) as~\cite{Kovtun:2012rj}
\begin{subequations} \label{eq:metric_fluctuation_components}
\begin{align}
    t_{\m\n} &=\le(\Pi_{\m\l} \Pi_{\n\s} - \frac{1}{d-1} \Pi_{\m\n} \Pi_{\l\s}\ri) h^{\l\s},
    \\
    u_\m &= \frac{1}{\Box} \Pi_{\m\nu} \p_\l h^{\nu\l},
    \\
    H &= \frac{1}{d-1} \le(d \frac{\p^\m \p^\n}{\Box} h_{\m\n}- h^\m{}_\m \ri),
    \\
    h &= \frac{1}{d-1} \le( h^\m{}_\m - \frac{\p^\m \p^\n}{\Box} h_{\m\n} \ri).
\end{align}
\end{subequations}

Under an infinitesimal diffeomorphism, \(x^m \to x^m + \xi^m\),  the scalar fields and metric transform as
\begin{equation}
    \F^a \to \F^a + \xi^m \p_m \F^a + \dots \; ,
    \qquad
    g_{mn} \to g_{mn} + \nabla_m \xi_n + \nabla_n \xi_m + \dots \; ,
\end{equation}
where the ellipses denote terms of higher order in \(\xi\). Applying the projections in Eq.~\eqref{eq:metric_fluctuation_components}, we can then extract the following gauge transformations for the fluctuation variables: 
\begin{gather}
    \d \psi^a = {\F^a}' \xi^\r,
    \qquad
    \d \n = \p_\r\xi^\r
    \qquad
    \d \n_\m = \h_{\m\n} e^{2B}  \p_\r \xi^\n + \p_\m \xi^\r,
    \nonumber
    \\
    \d t_{\m\n} = 0,
    \qquad
    \d u_\m = \Pi_{\m\n} \xi^\n,
    \qquad
    \d H = 2 \p_\m \xi^\m,
    \qquad
    \d h = 2B' \xi^\r.
\end{gather}
Using these transformations we identify gauge-invariant combinations of the fluctuation variables. With \(n\) scalar fields there are \((n+1)\)  combinations that do not include any radial derivatives of the fluctuations,
\begin{equation} \label{eq:gauge_invariant_variables_dynamical}
    \mathfrak{a}^a = \psi^a - \frac{{\F^a}'}{2 B'} h,
    \qquad
    \mathfrak{e}_{\m\n} = t_{\m\n}.
\end{equation}
In addition, there are three gauge-invariant combinations that we can form involving radial derivatives of the fluctuations,
\begin{equation}
    \mathfrak{b} = \n -  \le( \frac{h}{2B'}\ri)'
    \qquad
    \mathfrak{c} = \p^\m \n_\m - \frac{\Box h}{2B'} - \frac{e^{2B}}{2}H',
    \label{eq:gauge_invariant_variables_not_dynamical}
    \qquad
    \mathfrak{d}_\m = \Pi_\m{}^\lambda \n_\lambda - e^{2B} u_\m'.
\end{equation}
Notice that \(\mathfrak{c}\) and \(\mathfrak{d}_\m\) are essentially built from the longitudinal and transverse components of \(\n_\m\), respectively.

The action in Eq.~\eqref{eq:fluctuation_equations_action} leads to a cumbersome set of linearized equations of motion for the fluctuations. We use Eqs~\eqref{eq:gauge_invariant_variables_dynamical} and~\eqref{eq:gauge_invariant_variables_not_dynamical} to eliminate \((\psi^a,t_{\m\n},\n,\n_\m)\) from these linearized equations of motion in favor of the gauge invariant variables. One finds that the remaining gauge-dependent variables \((u_\m,h,H)\) then also disappear from the equations of motion---they represent residual gauge degrees of freedom. The fluctuations, written in the form of Eq.~\eqref{eq:gauge_invariant_variables_dynamical}, obey decoupled, second-order ordinary differential equations. Fourier transforming, so that \(\mathfrak{a}^a(\r,x^\m) = e^{i p_\m x^\m \mathfrak{a}^a(\r;p_\m)}\), the Fourier modes of the scalar fluctuations, \(\mathfrak{a}^a\), obey
\begin{equation} \label{eq:spin0_eom}
    e^{-dB} \le(e^{dB}{\mathfrak{a}^a}'\ri)' 
    + m^2 e^{-2B} \mathfrak{a}^a 
    - \le( \frac{ {V{\F^a}' {\F_b'}} }{(d-1)^2 B'^2} 
    + \frac{V^a {{\F_b'} + {\F^a}'} V_b }{(d-1) B'} 
     + V_{ab} \ri)\mathfrak{a}^b = 0,
\end{equation}
where \(V_a = \p V/\p \F^a\), \(V_{ab} = \p^2V/ \p \F^a \p \F^b\), and \(m^2 = - p_\m p^\m\). 
Recall that indices are raised and lowered with the identity matrix, $\delta_{ab}$. Similarly, the independent components of the gauge-invariant, spin-two fluctuation, \(\mathfrak{e}_{\m\n}\), obey the Klein--Gordon equation on the background metric,
\begin{equation} \label{eq:spin2_eom}
   e^{-d B} \le(e^{d B} \mathfrak{e}_{\m\n}'\ri)' +  m^2 e^{-2B}\mathfrak{e}_{\m\n} = 0.
\end{equation}
Once these equations are solved, the remaining gauge-invariant fluctuations are fixed by algebraic equations,
\begin{equation} \label{eq:algebraic_eom}
   \mathfrak{b} = \frac{\F_a' \mathfrak{a}^a}{2(d-1)B'},
    \qquad
  e^{-2B} \mathfrak{c} =- \frac{\F_a' {\mathfrak{a}^a}'}{2(d-1) B'}  + \frac{ V \F_a' \mathfrak{a}^a }{2(d-1)^2 B'^2}+ \frac{V_a \mathfrak{a}^a}{2(d-1) B'} ,
   \qquad
       \mathfrak{d}_\m  =0.
\end{equation}
The full set of fluctuation equations are completely solved by any \((\mathfrak{a}_a,\mathfrak{b},\mathfrak{c},\mathfrak{d}_\m,\mathfrak{e}_{\m\n})\) satisfying equations~\eqref{eq:spin0_eom},~\eqref{eq:spin2_eom}, and~\eqref{eq:algebraic_eom}.

Equations~\eqref{eq:spin2} and~\eqref{eq:spin0pertSimplified1}  in Sec.~\ref{Sec:B}  are recovered by setting \(d=3\) in equations~\eqref{eq:spin2_eom} and~\eqref{eq:spin0_eom}, respectively. To recover the notation used in Sec.~\ref{sec:fixed_point_annihilation}, we set \(d=4\), \(\r = L \log (z/z_0)\), and \(e^{2B} = L^2 f(z)/z^2\). Section~\ref{sec:fixed_point_annihilation} has a single scalar field, \(\F = \f\).

\section{Scalar fluctuations in the probe approximation} \label{sec:app_probe_fluctuations}

In this appendix, we discuss the details of the computation of the scalar fluctuations of model B when the fluctuations of the metric are ignored, which we referred to as the probe approximation. The motivation is to support our claim that the mode that becomes light is indeed a dilaton. In this approximation, the fluctuation equations become~\cite{Elander:2020csd}
\begin{equation}\label{eq:spin0pertSimplified}
       0=\left[\partial_{\rho}^2 + 3B' \partial_{\rho} + e^{-2B} m^2 \right] {\tilde{\mathfrak{a}}^a} -{V^a}_{b} {\tilde{\mathfrak{a}}^b}\,.
\end{equation}
Here, the ` $\tilde{\, }$ '  reminds us that these are not the actual perturbations, but their probe approximation.

Interestingly, the UV expansion of the perturbations are significantly altered. We find that
\begin{equation}\label{eq:UVprobeapprox}
\begin{aligned}
		{\tilde{\mathfrak{a}}^1}  &= 2\sqrt{3}\left( \vone +  \sone  \log(z)    +\frac{1}{4} \ps \stwo \log(z)^2 \right) z^ 2 +  O\left(z^4\right)\,,\\
		{\tilde{\mathfrak{a}}^2} &= \stwo z +
		 \Biggl[\vtwo+\frac{1}{24}\left(\stwo \phi _s^2-36 \phi _s (\sone-2
   \vone)- 12 m^2 \stwo \right) \log (z) \\
   &\qquad\qquad\qquad-\frac{3}{8} \phi _s  \left(\stwo \phi _s-4 \sone\right)\log ^2(z)+\frac{1}{4}
   \stwo \phi _s^2 \log ^3(z)\Biggr] z^3 + \OO(z^ 4)\,.\\
	\end{aligned}
\end{equation}
Notice the appearance of the coefficient $\stwo$ in both expansions. Fortunately, if we consider the combinations
\begin{equation}
    \psi_1(z) = \frac{1}{\log(z)}\left(\frac{\tilde{\mathfrak{a}}^1(z)}{2\sqrt{3}z^2} - \frac{1}{4}\tilde{\mathfrak{a}}^2(z)\phi(z)\frac{\log(z)}{z^2}\right)\,, \qquad \psi_2(z) = \frac{\tilde{\mathfrak{a}}_2}{z},
\end{equation}
we find that near the boundary
\begin{equation}\begin{aligned}
    \psi_1 &= \sone + \frac{\vone}{\log(z)} + O(z^2)\,,\\
   \psi_2 &= \stwo +
		 \Biggl[\vtwo+\frac{1}{24}\left(\stwo \phi _s^2-36 \phi _s (\sone-2
   \vone)- 12 m^2 \stwo\right) \log (z) \\
   &\qquad\qquad\qquad-\frac{3}{8} \phi _s  \left(\stwo \phi _s-4 \sone\right)\log ^2(z)+\frac{1}{4}
   \stwo \phi _s^2 \log ^3(z)\Biggr] z^2 + \OO(z^3)\,.\\
   \end{aligned}
\end{equation}
Thus, we identify $\stwo$ and $\sone$ as the coefficients that have to vanish for the solutions corresponding to the spectrum. Similarly, the perturbations near the  (IR) end of space read as follows:
\begin{equation}
    \label{}
    \begin{aligned}
        \tilde{\mathfrak{a}}^1&=\cone-\frac{(z-\zs)^2 \left(-2 \cone \gs L^2 V_B(\pIR)+6 \cone m^ 2
   \zs^2+\ctwo \gs L^2 V_B'(\pIR)\right)}{24 \left(\gs
   \zs^2\right)}+O\left((z-\zs)^3\right)\,,\\
   \tilde{\mathfrak{a}}^2&=\ctwo-\frac{(z-\zs)^2 \left(2 \cone \gs L^2 V_B'(\pIR)-\ctwo \gs L^2
   V_B''(\pIR)+\ctwo m^ 2 \zs^2\right)}{4 \left(\gs
   \zs^2\right)}+O\left((z-\zs)^3\right)\,,
    \end{aligned}
\end{equation}
with the appearance of  two undetermined parameters, $\cone$ and $\ctwo$.

The appearance of logarithmic divergences in Eq.~\eqref{eq:UVprobeapprox} makes the integration from the IR all the way up to the boundary challenging. In particular, it is complicated to proceed as we did with the full perturbations. There, we were looking for a combination of numerical solutions obtained from initial conditions at the IR for which the sources vanish at the UV. Here, instead, we  must compare numerical solutions obtained by evolving the differential equations from both sides, and merging them  at an intermediate point.

We proceed as follows. On the one hand, we set the sources to zero at the UV and make the choices $( \vone, \vtwo ) = (y_1,\pm y_2)$, with $y_i\in\mathds{R}$ positive real values, obtaining two different solutions $\{\jp_{(1,\text{\tiny I})}(z),\jp_{(2,\text{\tiny I})}(z)\}$ and $\{\jp_{(1,\text{\tiny II})}(z),\jp_{(2,\text{\tiny II})}(z)\}$. We find these solutions (numerically) and evolve them between the boundary and an intermediate matching point, $z = z_m\in(0,\zs)$. On the other hand, we solve the equations from the IR with $( \cone, \ctwo ) = (w_1,\pm w_2)$, with $w_i\in \mathds{R}$. This gives two additional numerical solutions, which we denote by $\{\jp_{(1,\text{\tiny III})}(z),\jp_{(2,\text{\tiny III})}(z)\}$ and $\{\jp_{(1,\text{\tiny IV})}(z),\jp_{(2,\text{\tiny IV})}(z)\}$. These are defined between the matching point $z_m$ and the IR $\zs$. 

With these two sets of solutions, we construct the matrix
\begin{equation}
M(m^2,z_m) = \left(\begin{array}{cccc}
	\jp_{(1,\text{\tiny I})}(z_m)\, &\jp_{(1,\text{\tiny I})}'(z_m)\, & \jp_{(2,\text{\tiny I})}(z_m)\, &\jp_{(2,\text{\tiny I})}'(z_m)\\
	\jp_{(1,\text{\tiny II})}(z_m)\, &\jp_{(1,\text{\tiny II})}'(z_m)\, & \jp_{(2,\text{\tiny II})}(z_m)\, &\jp_{(2,\text{\tiny II})}'(z_m)\\
	\jp_{(1,\text{\tiny III})}(z_m)\, &\jp_{(1,\text{\tiny III})}'(z_m)\, & \jp_{(2,\text{\tiny III})}(z_m)\, &\jp_{(2,\text{\tiny III})}'(z_m)\\
	\jp_{(1,\text{\tiny IV})}(z_m)\, &\jp_{(1,\text{\tiny IV})}'(z_m)\, & \jp_{(2,\text{\tiny IV})}(z_m)\, &\jp_{(2,\text{\tiny IV})}'(z_m)\,
\end{array}\right)\,.\end{equation}
If $\det M\neq 0$, it means that there is no linear combination of the IR solutions that reproduces the UV solutions. Consequently, modes with the sources $\sone = \stwo = 0$ cannot be connected to the regular IR. In contrast, if $\det M\neq 0$ there is a linear combination of $\text{I}$ and $\text{II}$ the value and derivative of which at $z_m$ coincide with a linear combination of $\text{III}$ and $\text{IV}$. 

Within this mid-point determinant method, the mass spectrum is given by those values of $m^2$ for which $\det(M(m^2,z_m)) = 0$. This set of $m^2$ should not depend on $z_m$, which we checked explicitly in the case of the numerical study presented in the body of the paper, and  shown in Fig.\ref{fig:fig_probe_fluctuations}.

\bibliographystyle{JHEP}
\bibliography{light-dilaton-cCFTs}
\end{document}